\documentclass[floatfix,superscriptaddress,showpacs,amssymb,10pt,aps,prd,reprint,longbibliography]{revtex4-1}

\usepackage{graphicx,epsfig,amssymb} 
\usepackage{amsmath,amsfonts, times}
\usepackage{bm} 

\usepackage[linktocpage,colorlinks]{hyperref}
\usepackage[caption=false]{subfig}
\usepackage[usenames]{color}     
\usepackage{natbib}
\usepackage{soul}
\usepackage[utf8x]{inputenc}
\usepackage{float}
\definecolor{coolblack}{rgb}{0.0, 0.18, 0.39}
\definecolor{darkred}{rgb}{0.5,0,0}
\definecolor{darkgreen}{rgb}{0,0.5,0}
\definecolor{darkblue}{rgb}{0,0,0.5}
\definecolor{lapislazuli}{rgb}{0.15, 0.38, 0.61}
\definecolor{venetianred}{rgb}{0.78, 0.03, 0.08}
\definecolor{bleudefrance}{rgb}{0.19, 0.55, 0.91}
\definecolor{dogwoodrose}{rgb}{0.84, 0.09, 0.41}
\hypersetup{colorlinks=true, citecolor=darkgreen, linkcolor=darkblue, 
	urlcolor = blue}

\newcommand{\dd}{\mathrm{d}}
\newcommand{\expe}{\operatorname{e}}
\newcommand{\ii}{\mathrm{i}}

\begin{document}
	
	\author{Qian Li}
	\address{School of Physics and Technology, Wuhan University, Wuhan, 430072, China}
	\author{Junji Jia}
	\email[Corresponding author:~]{junjijia@whu.edu.cn}
	\address{Department of Astronomy $\&$ MOE Key Laboratory of Artificial Micro- and Nano-structures, School of Physics and Technology, Wuhan University, Wuhan, 430072, China}
	\title{\Large  Matter around Schwarzschild black holes  in scalar-tensor theories: Absorption and Scattering}
	
	\begin{abstract}
	We investigate the absorption and scattering by a Schwarzschild black hole in scalar–tensor theories of gravity, where the coupling between matter and the scalar field induces different models for the effective mass of the scalar field. In model~I, a Bondi-type mass model described by the asymptotic mass $\mu_c$, horizon mass $\mu_H$, and profile slope $\lambda$, it is found that the absorption cross section increases with steeper $\lambda$, larger $\mu_c$ (especially at higher frequencies), or smaller $\mu_H$. The differential scattering cross section in this model shows the strongest dependence on the horizon mass $\mu_H$. When $\mu_H$ exceeds a critical value for a fixed incoming wave frequency $\omega$, no partial wave transmits into the black hole, flattening the differential scattering cross section as a function of angle before it increases again with further increase of $\mu_H$.
  Model~II, which considers a truncated accretion region outside some radius $r_0$, contains a potential well in its effective scattering potential. Its absorption cross section decreases in the low-frequency region as the accretion radius $r_0$ decreases, and more importantly, it shows resonance peaks at the quasibound wave frequencies due to resonances induced by the potential well. The differential scattering cross sections show dips around intermediate scattering angles when the parameters (mainly $\mu_H$ and $\omega$) are such that the resonantly scattered and non-resonant waves interfere destructively around these angles. In both models, absorption exhibits a zero-absorption band as $\omega$ approaches $\mu_c$ from above, and in both absorption and scattering, the effects of the parameters are found to diminish in the high-frequency limit.
	\end{abstract}
	
	
	\maketitle
	
	\section{Introduction}
Scalar–tensor theories, being among the simplest and most natural extensions of general relativity (GR), extend the gravitational sector beyond the metric tensor by incorporating one or more scalar fields nonminimally coupled to gravity. Such nonminimally coupled scalar fields can often be regarded as effective degrees of freedom emerging from quantum gravity theories in the low-energy limit \cite{Fujii:2003pa}. Therefore, the study of scalar–tensor theories not only helps to understand possible modifications of gravity but also provides a phenomenological framework that may offer insights into quantum gravity theories featuring additional scalar degrees of freedom. Furthermore, minimal or nonminimal couplings between gravity and scalar fields have been widely used in studies of potential models of dark energy, such as $f (R)$ theories of gravity \cite{DeFelice:2010aj}.

In scalar–tensor theories, black hole (BH) solutions generally differ from those in GR due to nontrivial scalar configurations, although in certain limits (e.g. constant field limit \cite{Sotiriou:2011dz}) they might reduce to the corresponding GR solutions. Consequently, the dynamics of perturbations and wave propagation in these spacetimes can also differ from those in GR, offering potential avenues for testing these gravity theories \cite{Barausse:2008xv}. Generically, if matter fields are present around the BH, the scalar field typically develops a nontrivial configuration, and the BH tends to acquire scalar “hair”.  However, Cardoso et al. \cite{Cardoso:2013fwa} have shown that there exists a subclass of theories in which the scalar field remains constant even in the presence of matter. In such cases, GR BHs remain solutions of the field equations, yet perturbations can destabilize the constant configuration, causing the scalar field to develop a nontrivial profile and endowing the BH with scalar hair. This nontrivial configuration arises because the coupling between the scalar field and the trace of the matter energy-momentum tensor can render a negative effective mass squared for the scalar field, leading to tachyonic instabilities that subsequently result in spontaneous scalarization \cite{Doneva:2022ewd}. Conversely, if the effective mass squared is positive, rapidly rotating BHs may experience superradiant instabilities \cite{Brito:2015oca}. Further studies \cite{Cardoso:2013opa} have indicated that the efficiency of superradiant instabilities depends on the matter distribution, BH spin, and the specific scalar–tensor theories considered.  Moreover, the effective scalar mass can give rise to long-lived modes, where scalar waves may undergo strong resonant amplification at particular frequencies. Based on these studies, Lingetti et al. \cite{Lingetti:2022psy} extended the analysis to Kerr BHs with arbitrary spin, incorporating realistic truncated accretion disk and coronal flow models to study superradiant instabilities, while Tanaka \cite{Tanaka:2025bfl} explored whether BHs surrounded by dark matter halos could develop scalar hair and superradiant instabilities.	

Similar resonant effects may also arise in other spacetimes, such as wormholes or BH remnants, where resonances can produce sharp and narrow features in the scalar wave absorption spectrum \cite{Lima:2020auu,Magalhaes:2023har,Furuta:2024jpw,Konoplya:2025hgp,Macedo:2018yoi,Delhom:2019btt}. Additionally, Lima Junior et al. studied the scattering of scalar waves in Simpson–Visser wormhole spacetimes and found that at resonant frequencies, the differential scattering cross section can be relatively suppressed \cite{LimaJunior:2022zvu}. These results suggest that in scalar–tensor theories, resonant effects of scalar waves may similarly influence BH absorption and scattering. Given that absorption and scattering represent one of the most compelling aspects of BH perturbation dynamics, numerous studies have examined the absorption and scattering of scalar fields by different BHs \cite{Sanchez:1977vz,Crispino:2009ki,Chen:2011jgd,Anacleto:2019tdj,Leite:2019eis,Anacleto:2020lel,Huang:2020bdf,Richarte:2021fbi,Wan:2022vcp,dePaula:2022kzz,Xavier:2023ljy,Li:2024xyu,Li:2025yoz,Baptista:2025ogh}.  Moreover, some studies further indicate that BH absorption and scattering of scalar waves are closely related to quasinormal modes (QNMs), whose decay rates provide direct information about the linear stability of the BH solution \cite{Decanini:2010fz,Folacci:2019cmc,Torres:2023nqg,OuldElHadj:2025hbl}. Therefore, investigating BH absorption and scattering in scalar–tensor theories may reveal deviations from GR predictions and provide potential observational signatures to distinguish between GR and scalar–tensor theories. Motivated by these considerations and as a preliminary exploration, this work focuses on Schwarzschild BHs, analyzing their absorption and scattering of scalar waves in scalar–tensor theories, as well as how these quantities vary with the effective matter distribution surrounding the BH.

The paper is organized as follows.  In Sec. \ref{sec:setup}, we briefly introduce the physical setup and two density profiles of effective mass. In Sec. \ref{sec:PWA}, we outline the partial wave method for the wave absorption and scattering and discuss the behavior of the effective potential. Sec. \ref{sec:results} is dedicated to giving the numerical results of wave absorption and scattering and analyzing the influence of the effective mass parameter on the cross sections. Finally, we summarize our main findings in Sec. \ref{sec:conclusion}. Throughout this work, the signature  $(-, \, +,\, +,\, +)$ for the spacetime metric and natural units $(G=c=\hbar=1)$ are adopted.

\section{Field equations and effective mass} 
\label{sec:setup}

In this section, we lay out the basic procedure for deriving the field equation and briefly discuss the choice of the effective mass. The derivation procedure is similar to that in Ref.~\cite{Cardoso:2013opa}.

\subsection{Background and perturbation equations} 
\label{sec:model}
A generic scalar-tensor theory of gravity  in the Jordan frame is described by the action
\begin{align}\label{eq:action}
	S=&\frac{1}{16\pi G}\int \dd^4x \sqrt{-g}\big(F(\phi)R-Z(\phi)g^{\mu\nu}\partial_{\mu}\phi\partial_{\nu}\phi-U(\phi)\big)\nonumber\\
	&+S_m(\Psi_m;g_{\mu\nu}),
\end{align}
where the Ricci scalar $R$ and the metric $g_{\mu\nu}$ characterize the spacetime geometry, while $\phi$ denotes a scalar field. The term $S_m$ represents the action of matter fields minimally coupled to the metric. The specific forms of the functions $F$, $Z$, and $U$ determine which scalar-tensor theories can be recovered. By applying a conformal transformation to the metric and redefining the scalar field
\begin{subequations}
	\begin{align}\label{eq:TF}
		g^E_{\mu\nu}&=F(\phi)g_{\mu\nu},\\
		\Phi(\phi)&=\frac{1}{\sqrt{4\pi}}\int d\phi\,\left[\frac{3}{4}\frac{F'(\phi)^2}{F(\phi)^2}+\frac{1}{2}\frac{Z(\phi)}{F(\phi)}\right]^{1/2},\\
		A(\Phi)&=F^{-1/2}(\phi),\\
		V(\Phi)&=\frac{U(\phi)}{F^2(\phi)},
	\end{align}
\end{subequations}
one can rewrite the action in the Einstein frame as 
\begin{align}\label{eq:Einstein}
	S=&\int \dd^4x \sqrt{-g^E}\left(\frac{R^E}{16\pi}-\frac{1}{2}g^E_{\mu\nu}\partial^\mu\Phi\partial^\nu\Phi-\frac{V(\Phi)}{16\pi}\right) \nonumber\\
	+&S_m(\Psi_m;A(\Phi)^2g_{\mu\nu}^E).	
\end{align}	

By varying $g^E_{\mu\nu}$ and $\Phi$, the field equations in the Einstein frame can be found as
\begin{align}\label{eq:fieldeq}
	G_{\mu\nu}^E=&8\pi 
	\left(T_{\mu\nu}^E+\partial_\mu\Phi\partial_\nu\Phi-\frac{g_{\mu\nu}^E
	}{2}(\partial\Phi)^2\right)-\frac{g_{\mu\nu}^E}{2} 
	V(\Phi),\\
	\square^E\Phi=&-\frac{A'(\Phi)}{A(\Phi)}T^E+\frac{V'(\Phi)}{16\pi}, \label{eq:fieldeq2}
\end{align}	
where the energy-momentum tensor in the Einstein frame is 
\begin{align}\label{eq: set}
	T^{E}_{\mu\nu}=\frac{-2}{\sqrt{-g^E}}\frac{\delta S_m(\Psi_m;A(\Phi)^2g_{\mu\nu}^E)}{\delta g_{\mu\nu}^E}=
	A^{2}(\Phi)T_{\mu\nu}
\end{align}
and $T_{\mu\nu}$ denotes the energy-momentum tensor in the Jordan frame. 
Let us assume that the potentials have a general analytic form around a constant value $\Phi_{0}$. Then the following expansions are valid
\begin{align}\label{eq:expand}
	V(\Phi)=\sum_{n=0}V_n(\Phi-\Phi_{0})^n,\\
	A(\Phi)=\sum_{n=0}A_n(\Phi-\Phi_{0})^n,
\end{align}	
where $V_n$ and $A_n$ are constant coefficients.
The first-order expansion of the field equations \eqref{eq:fieldeq} and \eqref{eq:fieldeq2} in terms of $\varphi\equiv\Phi-\Phi_{0}\ll1$ becomes
\begin{align} 
	&G_{\mu\nu}^E=8\pi T_{\mu\nu}^E-\frac{g_{\mu\nu}^E}{2} \left(V_0+V_1\right), \label{eq:Einstein_1}\\
	&\square^E\varphi=-\frac{A_1}{A_0}T^E+\frac{V_1}{16\pi} +\frac{V_2 \varphi}{8\pi}+\varphi T^E\left(\frac{A_1^2}{A_0^2}-2\frac{A_2}{A_0}\right), \label{eq:KG}
\end{align}
where $T^E$ and $T$ denote the traces of $T^E_{\mu\nu}$ and $T_{\mu\nu}$, respectively. In this work, we consider an asymptotically flat spacetime, which requires $V_{0} = V_{1} = 0$. Moreover, we also assume that $A_{1} = 0$, for which the field equations admit GR vacuum solutions with a constant scalar field (see Ref.~\cite{Cardoso:2013fwa} for more details). Therefore, the field equation for the scalar field simplifies to a Klein-Gordon (KG) equation
\begin{align} \label{eq:KG_1}	
	\left[\square^E-\mu^{2}(r)\right]\varphi=0,
\end{align}
with $\mu^{2}(r)=\frac{V_2}{8\pi}-\frac{2A_2}{A_0} T^E$. Clearly, the coupling between the scalar field and the matter fields in $T^E$ controls the effective mass squared of the scalar field. The quantity $\mu^{2}(r)$ can be positive or negative, depending on the signs and relative magnitudes of $V_2$, $A_0$, $A_2$, and $T^E$. When $\mu^{2}(r)$ is negative, it can give rise to spontaneous scalarization of the BH. When $\mu^{2}(r)$ is positive and the BH is rapidly rotating, a spontaneous superradiant instability can appear \cite{Cardoso:2013fwa}. In this work, we focus on the case in which $\mu^{2}(r)$ is positive.
	
\subsection{Effective Mass}\label{sec:mass}
In this subsection, we consider two different types of effective mass terms. Dima and Barausse \cite{Dima:2020rzg} proposed different types of effective mass terms in their study of plasma-driven superradiant instabilities. From their work, we select two models, namely model~I and model~II, for our analysis.

Model I adopts a pure power-law density profile to describe Bondi-type spherically symmetric accretion with a constant mass term $\mu_c$. The full profile is
\begin{align} \label{eq:model1}
	\mu^{2}{\text{I}}(r)=\mu^{2}{c}+ \mu^{2}{H}\left(\frac{r{h}}{r}\right)^{\lambda},
\end{align}
with $r_{h}$ being the event horizon radius, $\mu_c$ the asymptotic mass, and $\mu_H$ (when combined with $\mu_c$) characterizing the mass at the event horizon. $\lambda$ is the mass profile slope. Model II describes a truncated accretion sphere with the following profile
\begin{align} \label{eq:model2}
	\mu^{2}{\text{II}}(r)= \mu^{2}{c}+\mu^{2}{H}\Theta(r-r{0})\left(1-\frac{r_{0}}{r}\right) \left(\frac{r_{0}}{r}\right)^{\lambda}
\end{align}
with $\Theta(x)$ being the Heaviside step function and $r_0$ being the inner radius of the accretion region. The slope index $\lambda$ in both models should be positive so that the field have a  finite mass asymptotically.  Motivated by the work of Dima and Barausse \cite{Dima:2020rzg}, in this study we choose these parameters to be in the range $\mu_{c}=\{0.1,\,0.2,\,0.3,\,0.42,\,0.5\}~M^{-1}$, $\mu_{H}=\left[1-5\right] M^{-1}$, $\lambda=\{1,\,3/2,\,2\}$, and $r_{0}=\{0,\,3,\,6,\,8,\,10\}~M$, where $M$ is the BH mass. When $\mu_{H}=0$ or $r_{0}=0$, the analysis recovers the standard scenario of massive scalar waves scattering by a Schwarzschild BH. In the next section, we will analyze the effect of the above two mass profiles on the effective potential of scalar waves and show why we chose these two models.
	
\section{Partial wave analysis} \label{sec:PWA}
Sotiriou and Faraoni \cite{Sotiriou:2011dz} have shown that BHs in scalar-tensor theories of gravity can be described by the Kerr-Newman family of solutions when the backreaction of matter on the metric can be neglected. Therefore, for simplicity, we choose a Schwarzschild BH as the background metric for our study. To solve the KG equation \eqref{eq:KG_1}, the following ansatz is adopted
\begin{align}	\label{ansatz}
	\varphi_{\omega}= \sum_{l=0}^\infty\frac{R_{\omega l}(r)}{r}\expe^{-\ii \omega t} P_{l}(\cos\theta),
\end{align}
where $\omega$ is the frequency of the incident wave, $l$ is the angular quantum number, and $P_l$ are the Legendre polynomials. Substituting this ansatz into the KG equation \eqref{eq:KG_1} and separating variables, the radial equation is found to be
\begin{align}\label{eq:radial}
	\frac{\dd ^2}{\dd  r_{*}^2} R_{\omega l}(r)+\left[\omega^2-V_{\text{eff}}(r)\right] R_{\omega l}(r)=0,
\end{align}
where the effective potential is 
\begin{align}\label{eq:Veff}
	V_{\text{eff}}(r)=&f(r)\left(\mu^{2}(r)+ \frac{l(l+1)}{r^2}+\frac{f^\prime(r)}{r}
	\right),
\end{align}
with $f(r)=1-2M/r$ and $\mu^2(r)$ given in Eqs. \eqref{eq:model1} and \eqref{eq:model2}, and with $r_{*}$ being the tortoise coordinate linked to $r$ by $\dd r/\dd r_{*}=f(r)$.

\begin{figure*}[htp!]
	\centering
	\includegraphics[width=0.45\textwidth]{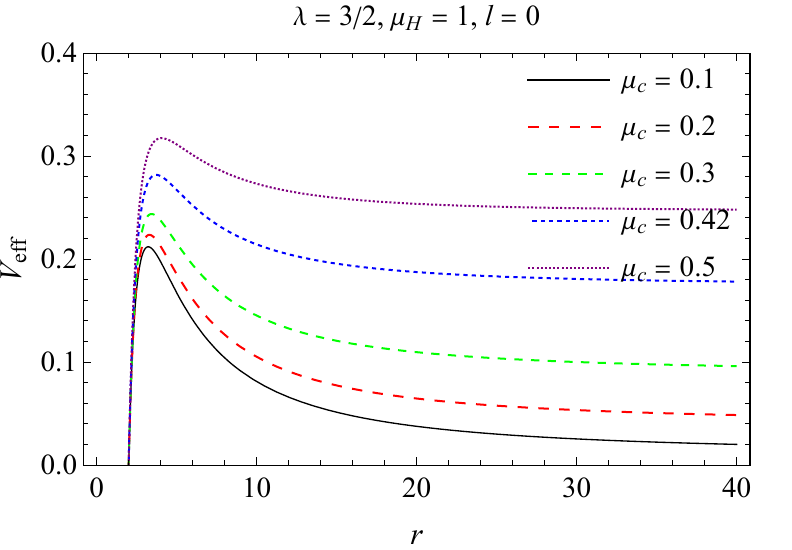}
	\includegraphics[width=0.45\textwidth]{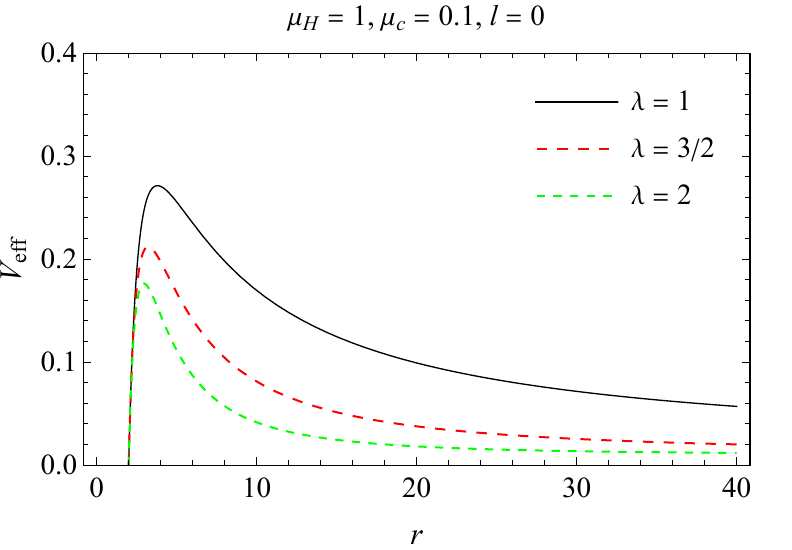}
	\\
	\includegraphics[width=0.45\textwidth]{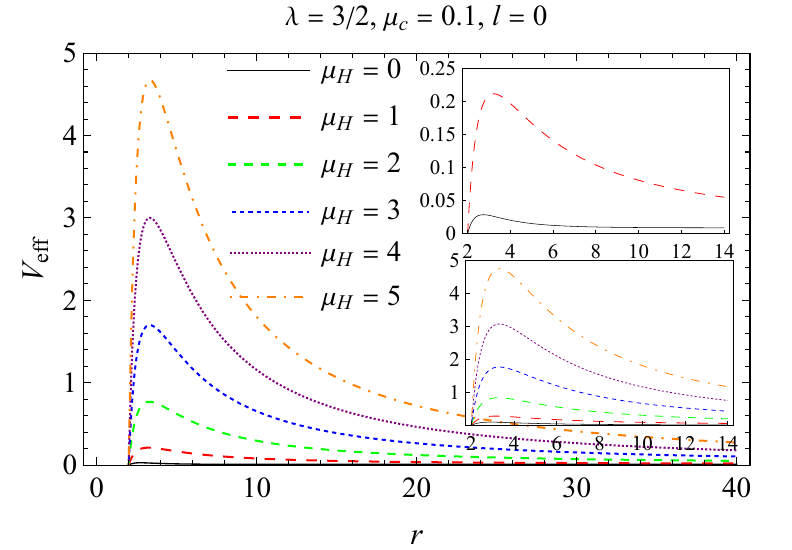}
	\includegraphics[width=0.45\textwidth]{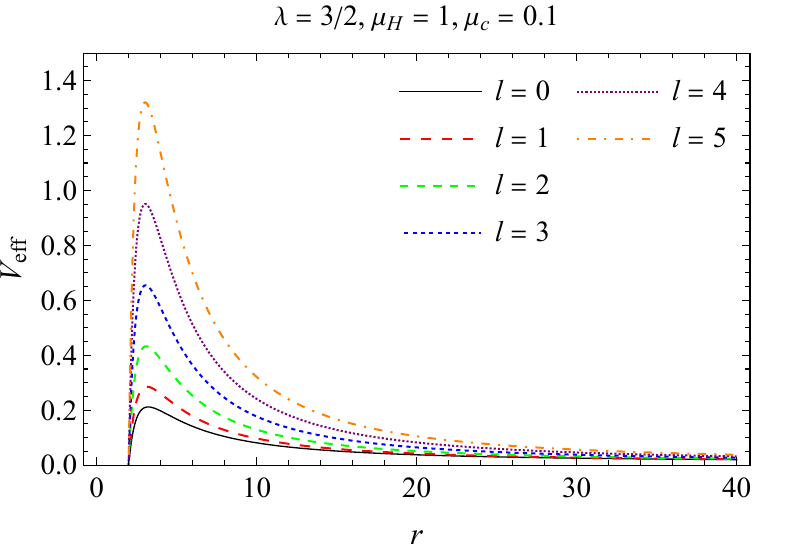}
	\caption{The effective potential in model~I as a function of $r$ for parameters $\mu_{c}$ (top left), $\lambda$ (top right), $\mu_{H}$ (bottom left), and $l$ (bottom right). The top inset shows $V_{\mathrm{eff}}$ for $l=0$ with $\mu_{H}=0$ and $\mu_{H}=1$. The bottom inset shows $V_{\mathrm{eff}}$ for $l=1$. }
	\label{fig:V1}
\end{figure*}

\begin{figure*}[htp!]
	\centering
	\includegraphics[width=0.45\textwidth]{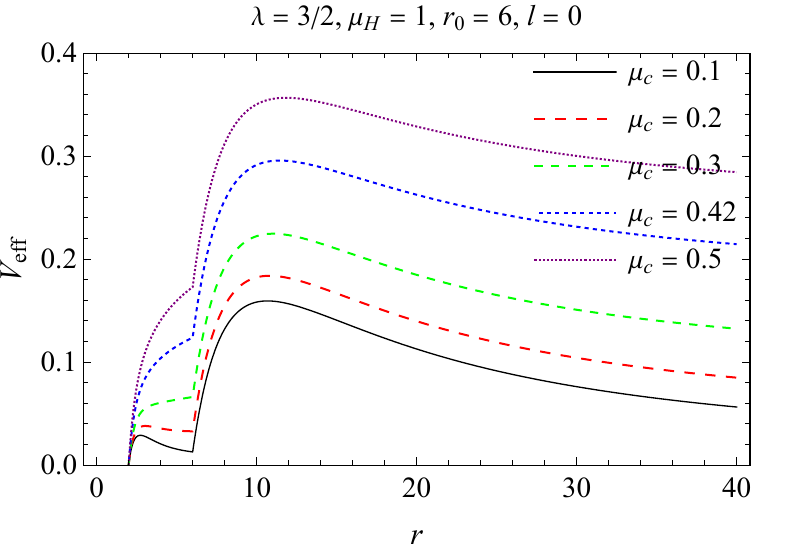}
	\includegraphics[width=0.45\textwidth]{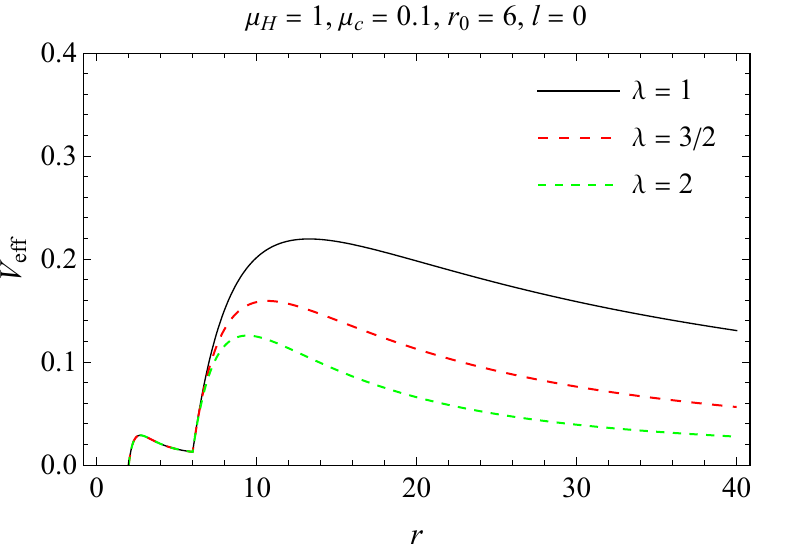}
	\\
	\includegraphics[width=0.45\textwidth]{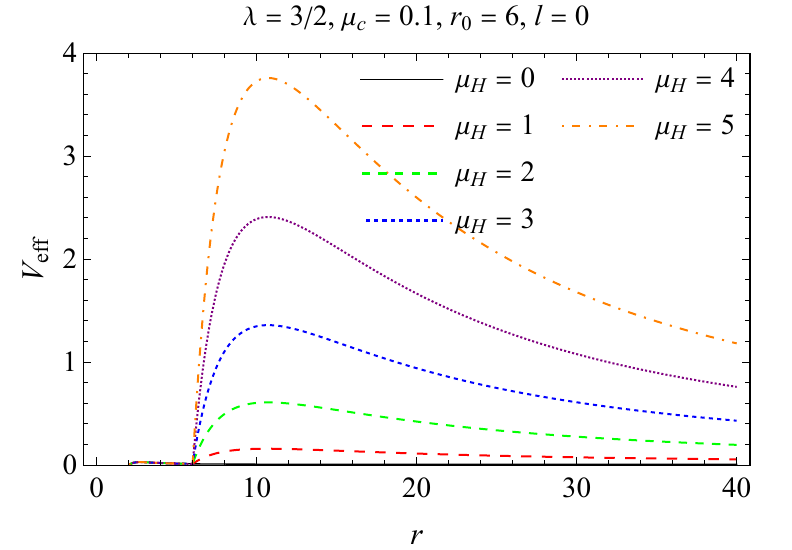}
	\includegraphics[width=0.45\textwidth]{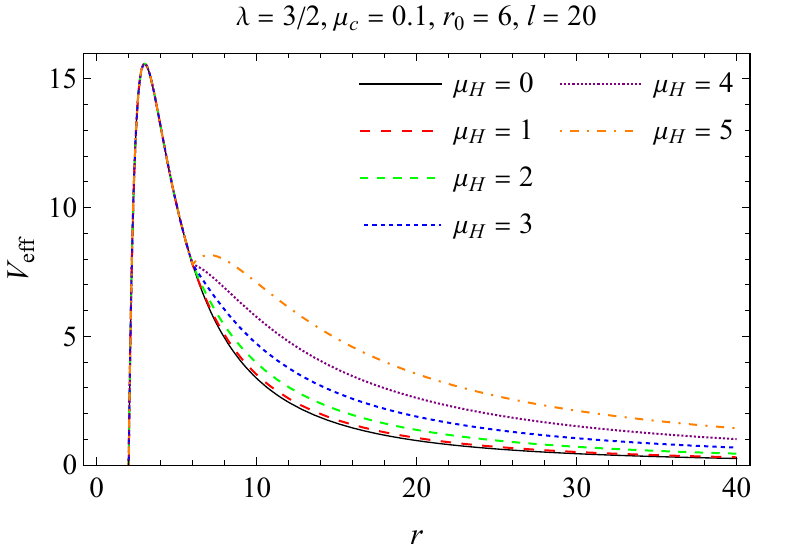}
	\\
	\includegraphics[width=0.45\textwidth]{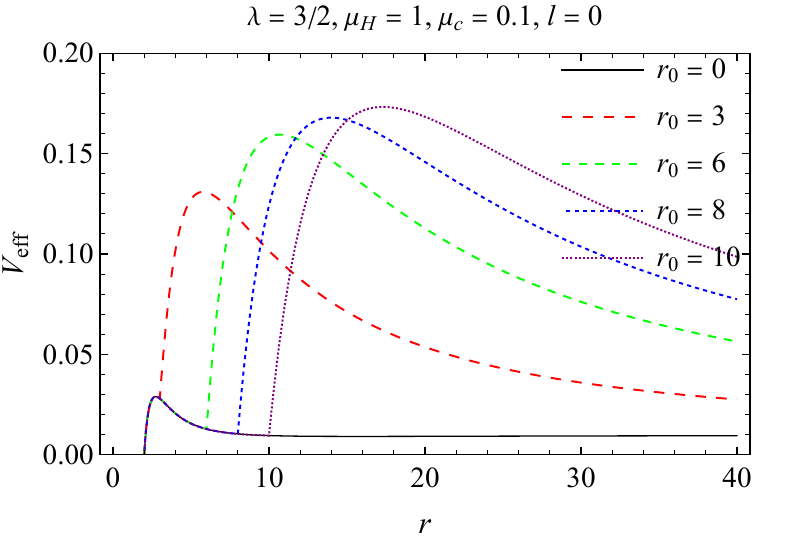}
	\includegraphics[width=0.45\textwidth]{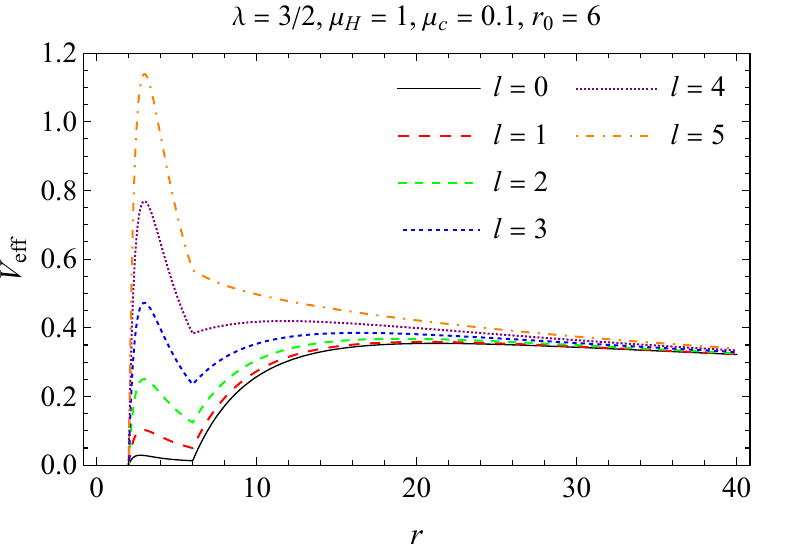}
	\caption{The effective potential in model~II  as a function of $r$ for parameters $\mu_{c}$ (top left), $\lambda$ (top right), $\mu_{H}$  (middle), $r_{0}$ (bottom left), and $l$ (bottom right).}
	\label{fig:V2}
\end{figure*}

In order to better understand the influence of the parameters in models I and II on wave absorption and scattering, which will be shown in Secs. \ref{sec:absorption} and \ref{sec:scattering}, we illustrate the effective potential \eqref{eq:Veff} in Figs. \ref{fig:V1} and \ref{fig:V2} for the pure power-law model \eqref{eq:model1} and the truncated accretion model \eqref{eq:model2}, respectively, as functions of $r$ for different values of the parameters $\mu_{c}$, $\lambda$, $\mu_{H}$ and/or $r_{0}$ (only for model~II), and $l$. Note that in the calculations henceforth, we set $M=1$, so the unit of $r$ is $M$, while the units of $\omega$ and $\mu$ are $M^{-1}$.
	
For model~I, it is seen from Fig. \ref{fig:V1} that the height of the effective potential increases with increasing asymptotic mass $\mu_{c}$ (top left), horizon mass $\mu_{H}$ (bottom left), and $l$ (bottom right). However, as shown in the top right panel of Fig. \ref{fig:V1}, the height of the potential barrier decreases with increasing slope index $\lambda$. Note that the enhancement of the potential is particularly dramatic for large values of $\mu_{H}$, compared to the effects of other parameters, in the ranges we considered. 
	
For model~II, where the truncation is included in the effective mass, we see from Fig. \ref{fig:V2} that the effective potential exhibits two distinct peaks for suitable parameters, indicating the emergence of an asymmetric potential well. Regarding the effect of parameters on the double-peak potential, we find from the top left panel that with the growth of the mass term $\mu_{c}$, the height of both inner and outer potentials increases, and the potential well gradually disappears. However, from the top right and middle left panels, it is observed that the slope index $\lambda$ and horizon mass $\mu_{H}$ only significantly affect the peak height of the outer potential, consistent with their effects in model~I. Note that in the middle left panel, the inner peaks are not apparent due to overlap and their small magnitude, but become very apparent for a large angular momentum number $l=20$ in the middle right panel. Clearly, for the inner peaks, both their locations and height remain unchanged for different $\mu_H$, although the outer peak, and therefore the total width of the barrier, becomes wider as $\mu_H$ increases.
The parameter $r_0$ is the position where the accretion disk is truncated, namely, the radius of the disk's inner edge. Thus, we observe from the bottom left plot that the larger the $r_0$, the wider and deeper the potential wells, and the higher the outer potential barrier, although the inner potential peaks remain unchanged. Finally, when we increase the angular momentum number $l$, we can see from the bottom right panel that although both the inner potential and outer potential increase in height, the inner potential increases faster, causing the potential well to gradually disappear.

Overall, the parameter dependence in model~I is relatively simple: the slope index $\lambda$ lowers the potential, whereas the remaining parameters enhance its height, without affecting the potential's structure. This behavior is analogous to the role of repulsive electromagnetic interactions in charged BHs \cite{dePaula:2024xnd}. By contrast, the introduction of $\Theta(x)$ in model~II fundamentally changes the structure of the potential, as the truncated accretion disk can generate a cavity around the BH and form a dynamically tunable potential well. In addition, here the depth and width of the well can be controlled by the parameters $\lambda$, $\mu_H$, and $r_0$, in sharp contrast to wormhole-induced wells whose structure is fixed by topology \cite{Lima:2020auu,LimaJunior:2022zvu}. Model I therefore allows parameter changes without altering the potential's shape, whereas model~II enables adjustable wells, revealing new mechanisms for accretion regulation. In this work, we directly study the consequences of such a potential well on the absorption and scattering processes.

For model~II above, it is known that for suitable parameters and angular momentum, the potential contains a well, which allows quasibound states of the scalar field. These quasibound states are long-lived QNMs with a small imaginary part, i.e., $\omega_{I}\ll \omega_{R}$ \cite{Berti:2009kk}. The boundary conditions of the QNMs are  
	\begin{align}\label{eq:qnm}
		R_{\omega l}(r) \approx
		\left\{
		\begin{array}{ll}
			\expe^{\ii \omega_\infty r_*},  &\mbox{for $r_*\rightarrow +\infty$},\\
			\expe^{-\ii \omega r_*},  &\mbox{for $r_*\rightarrow -\infty$},
		\end{array}
		\right.
	\end{align}
	with $\omega_\infty=\sqrt{\omega^2-\mu_{c}^2}$.
	
Although scalar-matter interactions in scalar-tensor gravity lead to an effective mass in the KG equation \eqref{eq:KG_1}, these scalar waves are still allowed to propagate to infinity, provided that $\omega$ is larger than the asymptotic mass $\mu_{c}$ \cite{Cardoso:2013opa}. Therefore, in this work, we  study the absorption and scattering of such waves by BHs. The asymptotic solutions of these waves satisfy the $in$
 modes condition \cite{Futterman:1988}, namely, 
\begin{align}\label{eq:solution}
	R_{\omega l}(r) \approx
		\left\{
		\begin{array}{ll}
			\expe^{-\ii\omega_\infty r_*} + \mathcal{R}_{\omega l} \expe^{\ii \omega_\infty r_*},  &\mbox{for $r_*\rightarrow +\infty$},\\
			\mathcal{T}_{\omega l} \, \expe^{-\ii \omega r_*},  &\mbox{for $r_*\rightarrow -\infty$},
		\end{array}
		\right.
	\end{align}
	where $\mathcal{R}_{\omega l}$ and $ \mathcal{T}_{\omega l}$ are the reflection and transmission coefficients, respectively, that satisfy the relation
	\begin{align}\label{eq:rtf}
		\left|\mathcal{R}_{\omega l}\right|^2 + \frac{\omega}{\omega_\infty} \left|\mathcal{T}_{\omega l}\right|^2 = 1.
	\end{align}

	Using the partial wave method, the total  absorption cross section of a massive scalar wave  incident on a spherically symmetric BH is written as  \cite{Sanchez:1977si,Benone:2014qaa}
	\begin{align} \label{eq:tacs}
		\sigma_{\mathrm{abs}} =\sum_{l=0}^{\infty}\sigma_l,
	\end{align}
	with  $\sigma_l$ being  the partial absorption cross section
	\begin{align} \label{eq:pacs}
		\sigma_l= \frac{\pi}{\omega_\infty^2}(2l+1)\left(1-\left|\mathcal{R}_{\omega l} \right|^2 \right).
	\end{align}
	In addition,  the differential scattering cross section can be expressed as
	\begin{align} \label{eq:dscs}
		\frac{\dd \sigma}{\dd \Omega} =\left|f(\theta)\right|^2,
	\end{align}
	with  scattering amplitude $f(\theta)$ being \cite{Sanchez:1977vz}
	\begin{align}\label{eq:sa}
		f(\theta) =\frac{1}{2\ii \omega_\infty}\sum_{l=0}^{\infty}(2l+1)\left[(-1)^{l+1} \mathcal{R}_{\omega l}-1\right]	P_{l}(\cos\theta).
	\end{align}

	In the following section, we numerically evaluate these cross sections using Eqs. \eqref{eq:tacs} and \eqref{eq:dscs}, and analyze the results.
	
	\section{Numerical results and analysis}\label{sec:results}
In this section, we present and discuss the numerical results for the absorption and scattering cross sections. The reflection coefficients $\mathcal{R}_{\omega l}$ are determined through numerical integration of the second-order differential equation \eqref{eq:radial} using a fourth-order Runge-Kutta algorithm (see Ref. \cite{Dolan:2012yc} for details of the calculation). Additionally, because Eq. \eqref{eq:sa} converges poorly when $\theta \approx 0$, we employ the reduced series technique as described in Ref. \cite{Yennie:1954zz} to mitigate this problem.
	
	\subsection{Absorption cross section}\label{sec:absorption}
	
	\begin{figure}[htp!]
		\centering
		\includegraphics[width=0.45\textwidth]{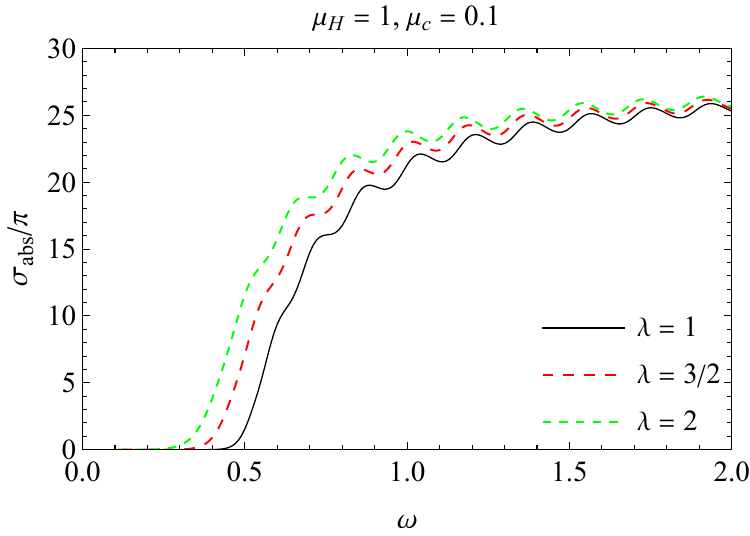}
		
		\includegraphics[width=0.45\textwidth]{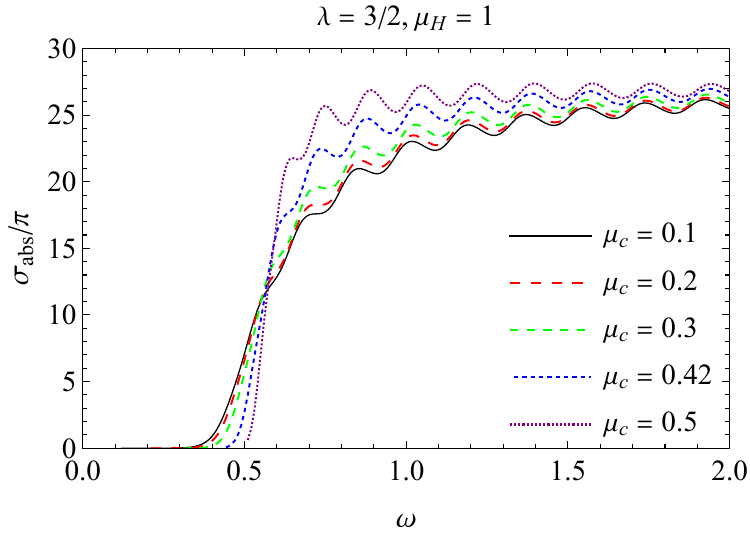}
		
		\includegraphics[width=0.45\textwidth]{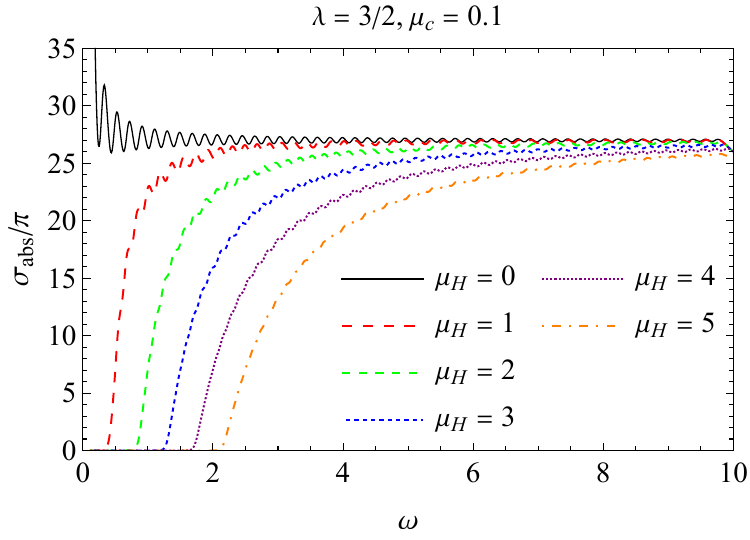}
		\caption{$\sigma_{\mathrm{abs}}$ in model~I as a function of $\omega$ for parameters $\lambda$ (top), $\mu_{c}$ (middle),  and $\mu_{H}$ (bottom).  }
		\label{fig:abs1}
	\end{figure}
	
In Fig. \ref{fig:abs1}, we show the total absorption cross section in model~I for different parameters of $\lambda$, $\mu_{c}$, and $\mu_{H}$. As can be seen from the top and middle panels, the total absorption cross section increases as the slope index $\lambda$ and asymptotic mass $\mu_c$ increase. For $\lambda$, this is because a steeper slope lowers the potential, thereby enhancing the transmission coefficient (see the top left panel of Fig. \ref{fig:V1}). However, for the asymptotic mass $\mu_c$, increasing it not only changes the transmission coefficient, but also increases the first factor $1/\omega_\infty^2$ in the partial absorption cross section \eqref{eq:pacs}. Therefore, when $\mu_c$ increases, despite the rise in the effective potential which would usually suppress absorption, the total absorption cross section still increases primarily in the high-frequency regime.
This behavior is consistent with the effect of the mass in the massive scalar field case \cite{Benone:2014qaa}. However, a significant difference between this work and previous literature \cite{Benone:2014qaa} occurs when $\omega \rightarrow \mu_c$, where the total absorption cross section approaches zero here for $\mu_H>0$, instead of diverging in the plain BH scattering of a massive scalar field, as shown in Ref. \cite{Benone:2014qaa}, or, similarly, as clearly seen in the $\mu_H=0$ line in the bottom plot.  The reason for this difference at lower frequency is that a nonzero horizon mass $\mu_H$ (as chosen in our case) introduces a much higher potential peak, which prevents the wave from being absorbed by the BH, particularly for low-$l$ partial waves. For instance, when $l=0$, the potential peak height rises from $\sim0.03$ for $\mu_H=0$ to $\sim0.21$ for $\mu_H=1$, a sevenfold increase (see the top inset of the bottom left panel of Fig. \ref{fig:V1}). Therefore, as $\mu_H$ increases, the total absorption cross section exhibits two interesting behaviors: (i) a monotonic decrease in overall magnitude, and (ii) the widening of a distinct zero-absorption band at low frequencies. This is similar to the case of a scalar field with the same sign of charge incident on a charged BH, where the repulsive interaction suppresses low-energy absorption \cite{Benone:2015bst,Li:2024xyu}; however, in our work no superradiance is induced.

\begin{figure*}[htp!]
	\centering
	\includegraphics[width=0.45\textwidth]{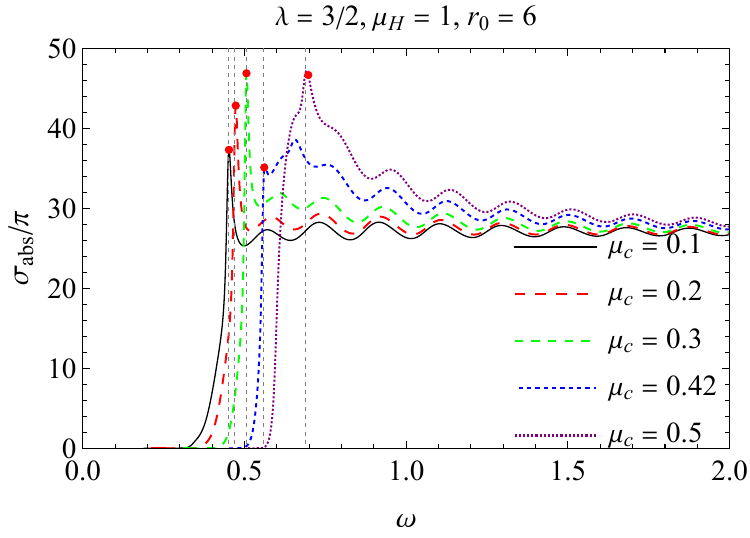}\includegraphics[width=0.45\textwidth]{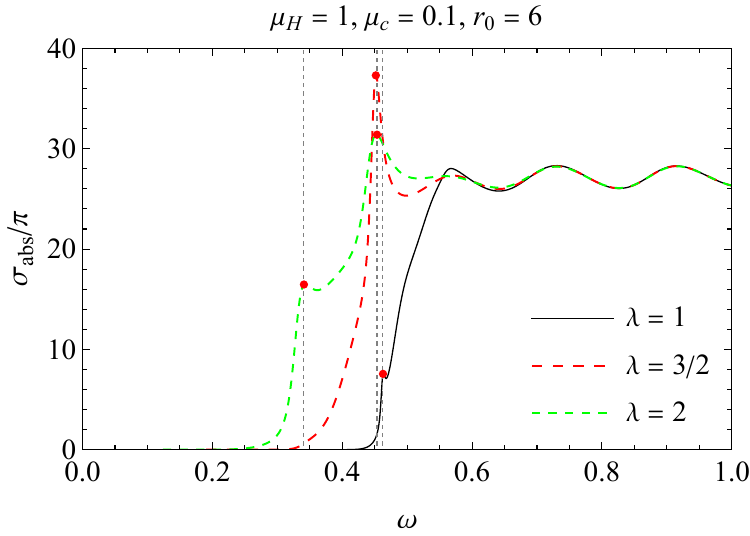}
	
	\includegraphics[width=0.45\textwidth]{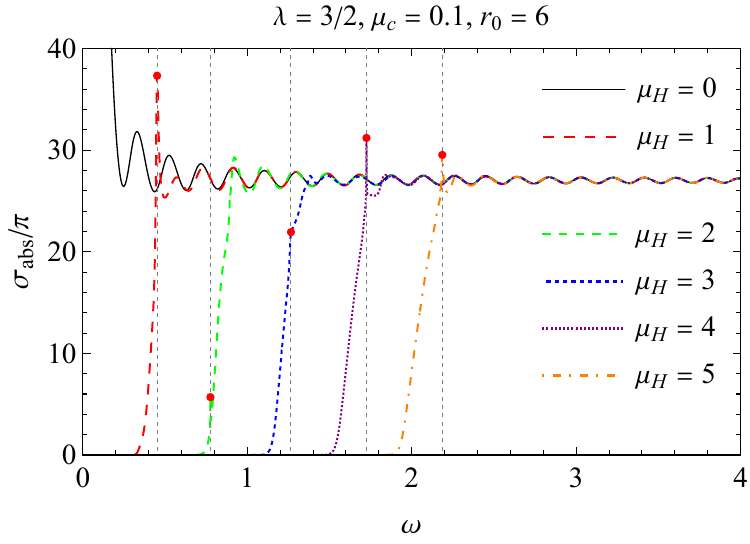}\includegraphics[width=0.45\textwidth]{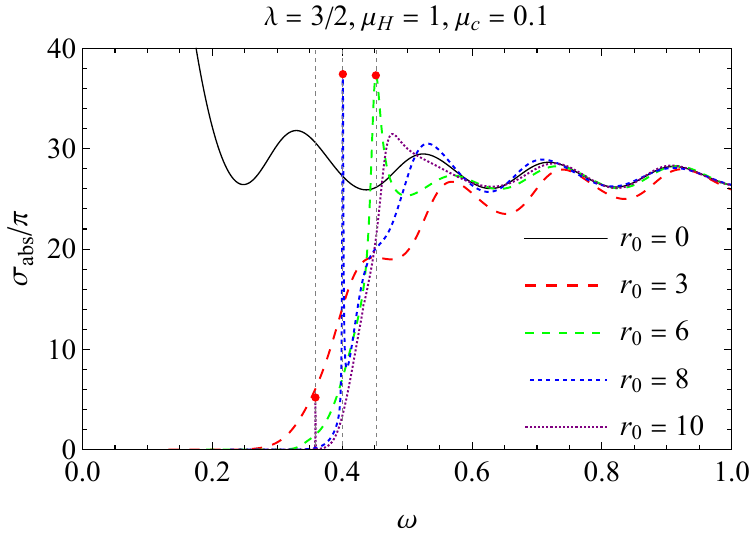}
	\caption{$\sigma_{\mathrm{abs}}$ in model~II as a function of $\omega$ for $\mu_{c}$ (top left), $\lambda$ (top right), $\mu_{H}$ (bottom left), and $r_{0}$ (bottom right). Red points and gray dashed lines mark the resonance peaks and frequencies, respectively. Note that the red point near $\omega=3.5$ belongs to the purple dotted curve. }
	\label{fig:abs2}
\end{figure*}
	
\begin{figure*}[htp!]
	\centering
	\includegraphics[width=0.45\textwidth]{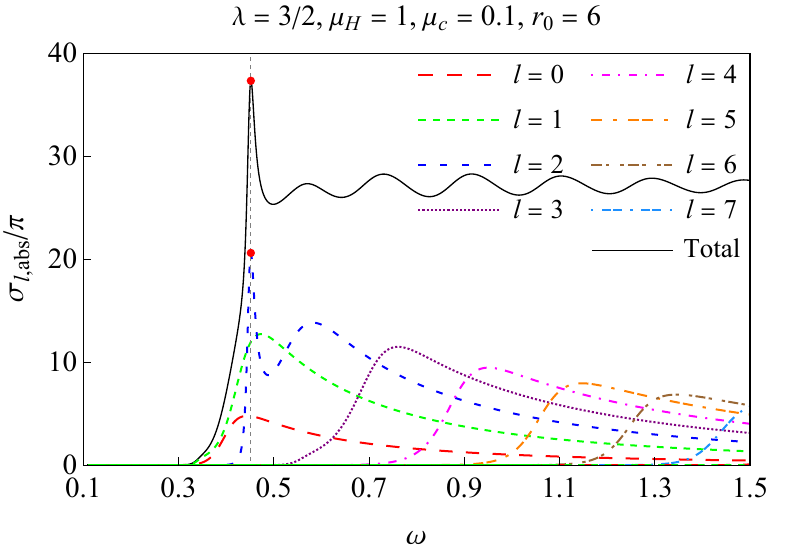}\includegraphics[width=0.45\textwidth]{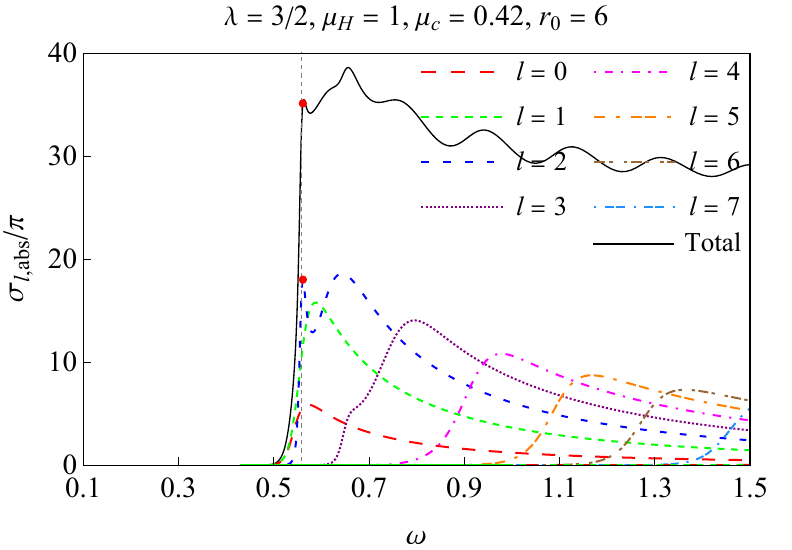}
	\includegraphics[width=0.45\textwidth]{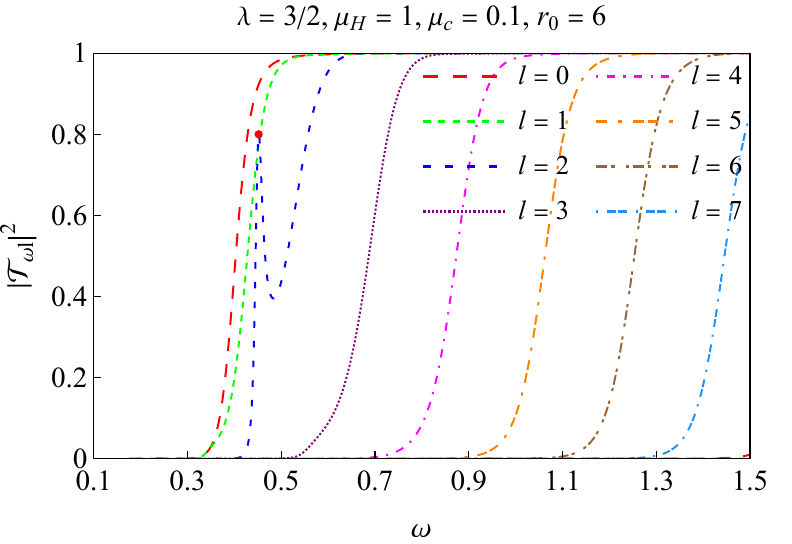}\includegraphics[width=0.45\textwidth]{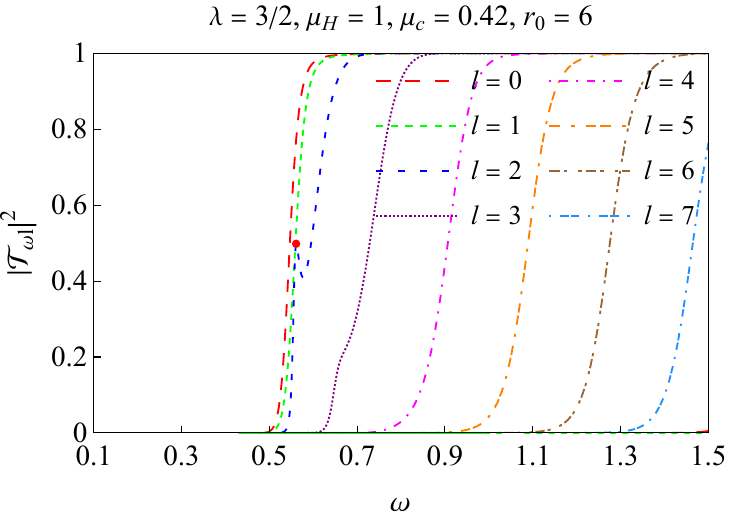}
	\includegraphics[width=0.45\textwidth]{V2_l.pdf}\includegraphics[width=0.45\textwidth]{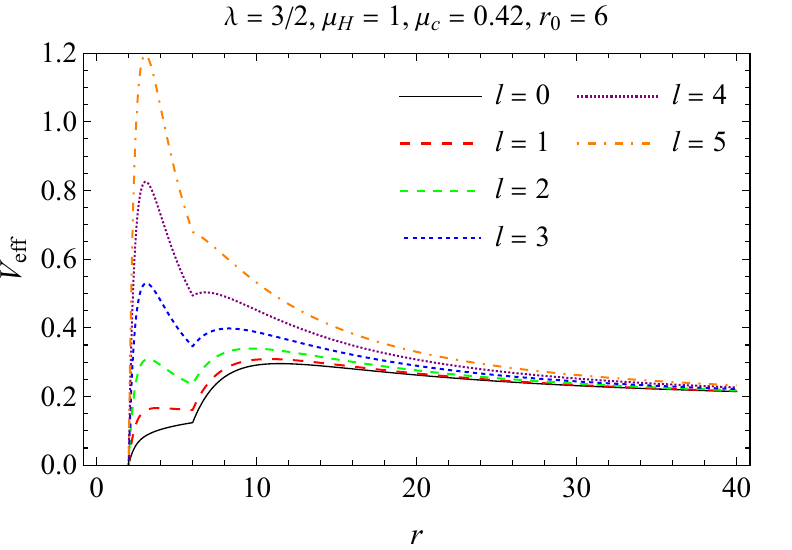}
	\caption{$\sigma_{l,\mathrm{abs}}$ (top), transmission coefficients (center), and effective potential (bottom) in model~II for the parameters $\mu_{c}=0.1$ (left) and $\mu_{c}=0.42$ (right). The red points mark the resonance positions at specific values of $l$ and $\omega$, while the vertical gray dashed lines indicate the real parts of the quasibound frequencies $\omega_{R}$ listed in Table~\ref{tab:qsb}.}
	\label{fig:abs3}
\end{figure*}

In Fig. \ref{fig:abs2}, we present the total absorption cross section in model~II for different parameters $\mu_{c}$, $\lambda$, $\mu_{H}$, and $r_0$. For partial waves that possess an appropriate potential well, a resonance occurs, exhibiting a Breit–Wigner profile with finite amplitude and an extremely narrow width \cite{Cardoso:2013opa}. Therefore, we mark both the values of the total absorption cross sections (red dots) and the real parts of the quasibound frequencies (vertical gray dashed lines), around which the resonances occur, in Fig. \ref{fig:abs2}. The corresponding quasibound frequencies for different parameters are tabulated in Table \ref{tab:qsb}, computed via the direct integration method \cite{Chandrasekhar:1975zza,Molina:2010fb} with the boundary conditions \eqref{eq:qnm}. Detailed calculation procedures can be found in Ref. \cite{Molina:2010fb}. It should be noted that here, we mark only the corresponding resonance peaks that become clearly distinguishable in the partial wave absorption (see~Fig.~\ref{fig:abs3}).

	\begin{table}[htp!]
		\caption{Quasibound frequencies for different values of $\mu_{c}$,  $\lambda$,  $\mu_{H}$ and $r_{0}$. The default parameters are set to  $\lambda=3/2$, $\mu_{c}=0.1$, $\mu_{H}=1$, $r_{0}=6$.}	\label{tab:qsb}
		\begin{tabular}{lcc}
			\hline\hline
			Parameter & Quasibound freq. ($\omega_{R}+\omega_{I}\ii$) & Angular num. ($l$) \\
			\hline
			$\mu_c=0.1$  & $0.4521676 - 0.0007954\ii$ & 2 \\
			$\mu_c=0.2$  & $0.4727889 - 0.0006170\ii$ & 2 \\
			$\mu_c=0.3$  & $0.5062011 - 0.0014030\ii$ & 2 \\
			$\mu_c=0.42$ & $0.5585553 - 0.0070512\ii$ & 2 \\
			$\mu_c=0.5$  & $0.6885291 - 0.0002066\ii$ & 3 \\
			\hline
			$\lambda=1$ & $0.4619803 - 0.0036621\ii$ & 2 \\
			$\lambda=3/2$ & $0.4521676 - 0.0007954\ii$ & 2 \\
			$\lambda=2$ & $0.4502344 - 0.0007478\ii$ & 2 \\
			$\lambda=2$ & $0.3398614 - 0.0006496\ii$ & 3 \\
			\hline
			$\mu_H=1$    & $0.4521676 - 0.0007954\ii$ & 2 \\
			$\mu_H=2$    & $0.7759294 - 0.0007373\ii$ & 4 \\
			$\mu_H=3$    & $1.2628087 - 0.0080437\ii$ & 6 \\
			$\mu_H=4$    & $1.7246731- 0.0008000\ii$ & 9 \\
			$\mu_H=5$    & $2.1852735 - 0.0012977\ii$ & 11 \\
			\hline
			$r_0=6$     & $0.4521676 - 0.0007954\ii$ & 2 \\
			$r_0=8$     & $0.4005709 - 0.0006154\ii$ & 2 \\
			$r_0=10$    & $0.3587535 - 0.0001305\ii$ & 2 \\
			\hline\hline
		\end{tabular}
	\end{table}
	
In the top left panel, we plot the absorption cross section $\sigma_{\mathrm{abs}}$ for several values of $\mu_c$. It is observed that $\sigma_{\mathrm{abs}}$ increases monotonically with $\mu_c$ for a fixed frequency in the comparable range ($\omega>0.8$), which is consistent with the result obtained in model~I in Fig. \ref{fig:abs1}. We see from the top right and bottom panels that the parameters $\lambda$, $\mu_{H}$, and $r_0$ significantly affect the absorption spectrum at low frequencies, while their influence on the absorption cross section in the high-frequency regime is almost imperceptible. For low frequency waves, the increase of the accretion radius $r_0$ decreases the absorption. The onset of the absorption is pushed towards smaller frequencies for a steeper index $\lambda$ and a smaller horizon mass $\mu_H$, similar to the effects of these two parameters in Fig. \ref{fig:abs1}. For the high-frequency behavior in these three cases, we observe that the absorption approaches its value for the $\mu_H=0$ or $r_0=0$ case, which in turn agrees with the other $\mu_H$ and $r_0$ cases at high frequencies. Since $\mu_H = 0$ or $r_0=0$ corresponds to the pure Schwarzschild BH scattering, this means that at high frequencies, the absorption cross section resembles the case of a massive scalar field scattered by a Schwarzschild BH. The effectiveness of $\lambda$, $\mu_H$, and $r_0$ is concentrated around low frequencies because these parameters modify only the outer potential peak at larger radii, as seen from Fig. \ref{fig:V2}.

As previously mentioned, the appearance and characteristics of the effective potential and, consequently, the resonance peaks depend on the parameters $(\mu_{c}, \lambda, \mu_{H}, r_0)$ and the angular momentum $l$. From Fig. \ref{fig:abs2}, we see that for our parameter choices, the resonance peaks in the absorption spectra are observed in all cases except for $\mu_H = 0$ and $r_0=0,3$. We notice from the top right panel that for $\lambda = 2$, two resonance peaks are observable. Moreover, it is worth noting that the resonance features vary across different cases, with some resonance peaks being sharp while others are barely recognizable. These differences can be plausibly traced back to the structure of the potential wells studied in Fig. \ref{fig:V2}. For the $\mu_H = 0$ and $r_0=0$ case, no resonance peaks appear because no potential well is formed. For the $r_0 = 3$ case, the potential well becomes extremely narrow and nearly vanishes entirely, thus failing to produce resonances, as shown in the bottom left panel of Fig. \ref{fig:V2}.

In Fig.~\ref{fig:abs3}, we take $\mu_{c}=0.1$ and $\mu_{c}=0.42$ as examples to explain, using partial absorption cross sections (top panels), transmission coefficients (middle panels), and the effective potentials (bottom panels), why some resonances in the total absorption spectrum appear as pronounced sharp peaks while others do not. For both of these two choices of $\mu_c$, we observe from the top and middle panels that only the $l=2$ partial wave shows resonance features for the chosen parameter settings. More importantly, the heights and sharpness of the peaks in the partial absorption cross section $\sigma_2$ as well as in the transmission coefficients $\mathcal{T}_{\omega 2}$ are different in these two cases. Using the Breit-Wigner expression for nuclear scattering \cite{Breit:1936zzb}, the transmission probability can be approximated
	\begin{align} \label{eq:BW}
		\left|\mathcal{T}_{\omega l}\right|^2 \propto \frac{1}{\left(\omega-\omega_{R}\right)^2+\omega_{I}^2}.
	\end{align}
Thus, it is clear that the height/sharpness of the resonance peaks is determined by the imaginary part $\omega_{I}$ of the quasibound frequency. The $\omega_I$, on the other hand, is usually related to the detailed properties of the effective potential, such as the width and depth of the potential well, its smoothness, and its general height. Sometimes the properties of the incident wave, such as its mass, are also relevant. In general, for wells that are otherwise similar, the deeper and wider the well, the stronger the resonance and the higher/sharper the resonance peaks in the transmission coefficient $\mathcal{T}_{\omega l}$ and partial absorption $\sigma_l$. 
From the effective potentials in the bottom panels in Fig.~\ref{fig:abs3}, a careful comparison between the $l=2$ potentials shows that the well for the $\mu_c=0.1$ case is slightly deeper than that for the $\mu_c=0.42$ case, while their widths are similar. Therefore, one would expect that the resonance for the former case should be stronger than that for the latter, which is exactly what is observed in the first two rows of Fig.~\ref{fig:abs3}. 
One more reason that the resonance peak in the case of $\mu_c=0.1$ is sharper than in the $\mu_c=0.42$ case is that, in the former case, the extra resonance accumulates positively on top of the peaks of the $l=0$ and $l=1$ partial absorption cross sections, as seen from the top left panel. For the latter case, however, the resonance contribution accumulates only on the much lower $\sigma_0$ peak.

Finally, let us remind that these resonance peaks enable the distinction between Schwarzschild BHs in GR and scalar-tensor theories through low-frequency absorption cross sections. Although wormholes \cite{Lima:2020auu,Magalhaes:2023har,Furuta:2024jpw,Konoplya:2025hgp}, extreme compact objects \cite{Macedo:2018yoi}, and BH remnants in metric-affine gravity \cite{Delhom:2019btt} all induce numerous significant resonance peaks in the absorption spectrum, the underlying mechanisms are fundamentally different. The potential well caused by the effective mass of the scalar field in our study exhibits obvious asymmetry and yields significantly fewer resonance peaks compared with these objects \cite{Lima:2020auu,Magalhaes:2023har,Furuta:2024jpw,Konoplya:2025hgp,Macedo:2018yoi,Delhom:2019btt}.

	\subsection{Scattering cross section}\label{sec:scattering}
	
	\begin{figure*}[htp!]
		\centering
		\includegraphics[width=0.33\textwidth]{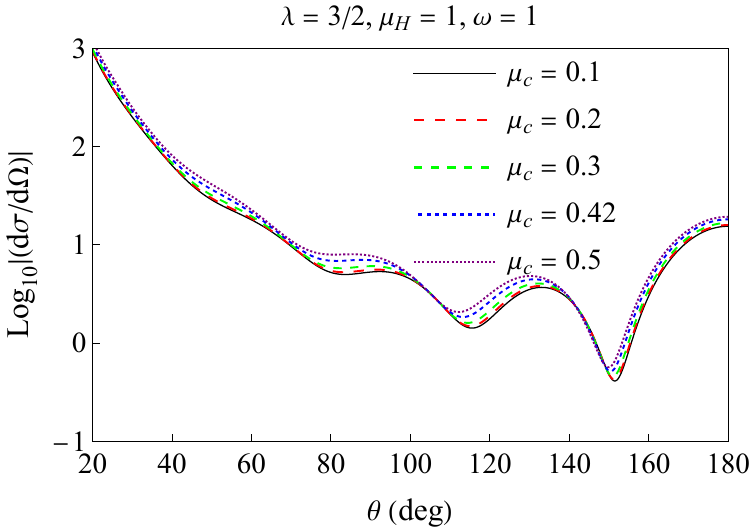}\includegraphics[width=0.33\textwidth]{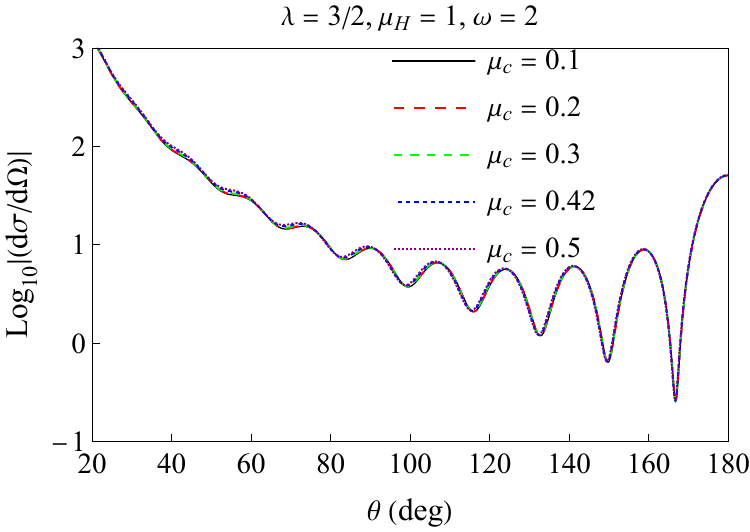}\includegraphics[width=0.33\textwidth]{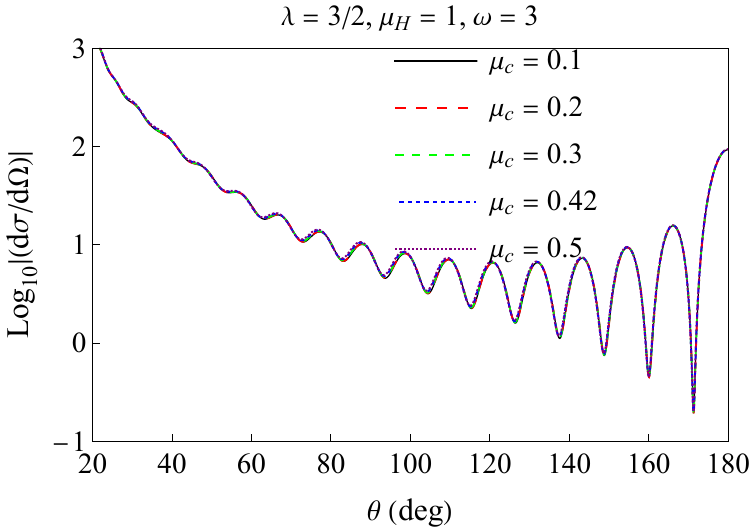}
		
		\includegraphics[width=0.33\textwidth]{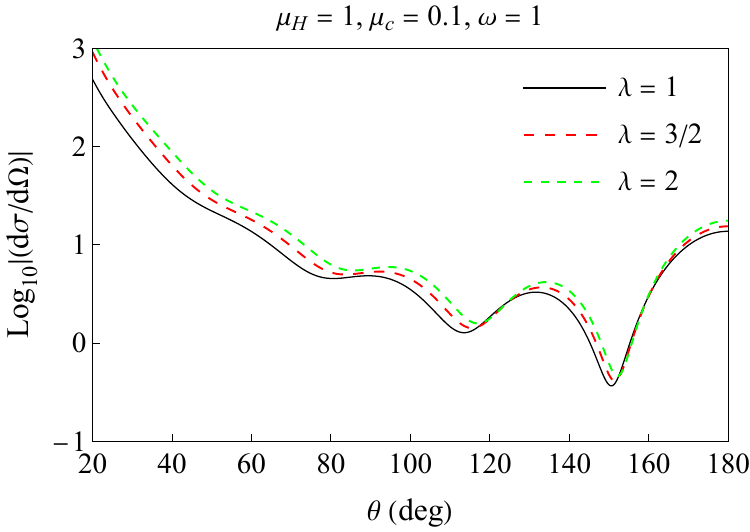}\includegraphics[width=0.33\textwidth]{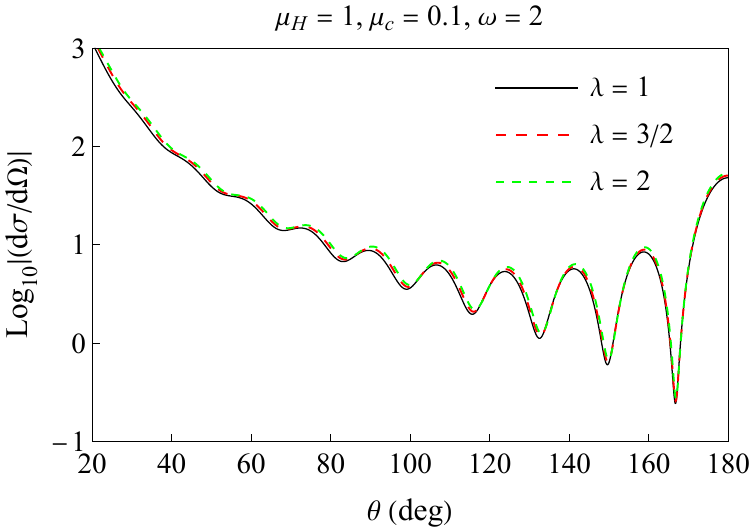}\includegraphics[width=0.33\textwidth]{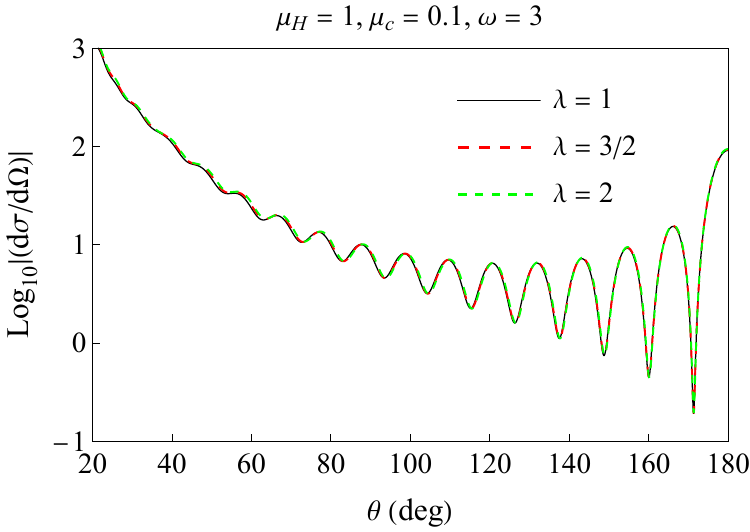}
		
		\includegraphics[width=0.33\textwidth]{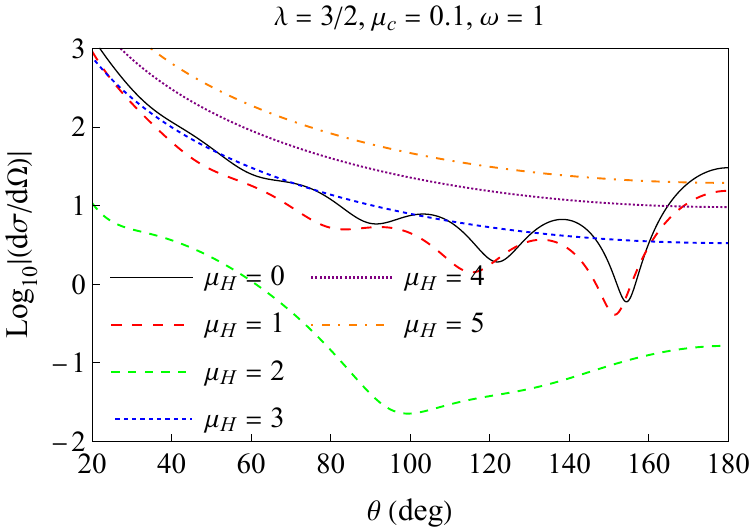}\includegraphics[width=0.33\textwidth]{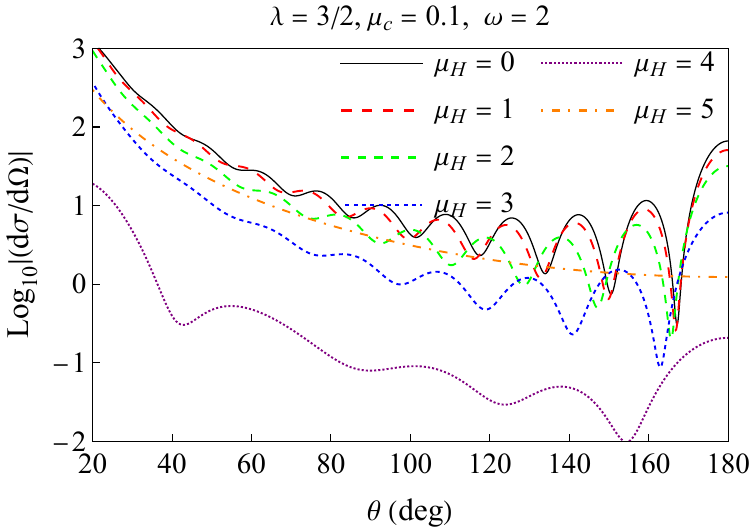}\includegraphics[width=0.33\textwidth]{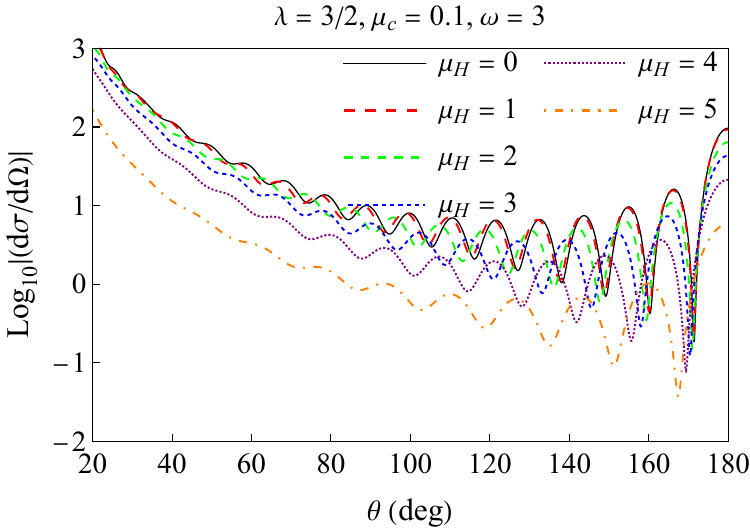}
		
		\caption{The differential scattering cross section in model~I for parameters $\mu_{c}$ (top), $\lambda$ (middle), and $\mu_{H}$ (bottom) at $\omega=1$ (left), $\omega=2$ (central), and $\omega=3$ (right).}
		\label{fig:sca1}
	\end{figure*}
	
In Fig. \ref{fig:sca1}, we show the differential scattering cross section in model~I for various values of the parameters $\mu_{c}$ (top row), $\lambda$ (middle row), and $\mu_{H}$ (bottom row) at the incident frequencies $\omega=1$ (left column), $\omega=2$ (central column), and $\omega=3$ (right column). Besides the common features in massive scalar field scattering \cite{Li:2024xyu}, such as the divergence as $\theta \to 0^\circ$, the glory scattering as $\theta \to 180^\circ$, as well as the increase of the oscillation amplitude as $\theta$ increases, we also observe unique effects of these parameters in the differential scattering cross sections.

For the effect of the asymptotic mass $\mu_c$, we observe that when the incident frequency is $\omega=1$, both the differential scattering cross section and the interference fringe width increase monotonically with $\mu_{c}$, as shown in the top left panel. This behavior agrees with results for the case of charged massive scalar fields \cite{Li:2025jpa} and can be explained by the factor $1/\omega_\infty^2 = 1/(\omega^2 - \mu_c^2)$ in the differential cross section in Eq. \eqref{eq:dscs}.  
For the slope index $\lambda$, we can see from the middle row that the larger the $\lambda$, the narrower the interference fringes and the larger the differential scattering cross section, especially at low incident frequencies. The influence of both $\mu_{c}$ and $\lambda$ on $\dd\sigma/\dd\Omega$ diminishes gradually as the incident frequency increases, becoming almost negligible for $\omega=3$ (see the top right and middle right panels). This effect arises from the fact that both $\mu_{c}$ and $\lambda$ predominantly influence the potential associated with low-$l$ partial waves or the potential shape (mainly the width) in the low-energy part, to which low-frequency waves are particularly sensitive.

For the effect of the horizon mass $\mu_H$, by carefully comparing with the bottom inset in the lower-left panel of Fig. \ref{fig:V1}, we observe that for each fixed frequency $\omega$ in the lowest row of Fig. \ref{fig:sca1}, as $\mu_H$ increases such that $\omega^2$ becomes lower than the maximum of the effective potential even for the $p$ ($l=1$) wave, the interference fringes of the differential cross section disappear. More precisely, for $\omega^2 = 1,\,4,\,9$, this happens between $\mu_H = 2,\,3$, $\mu_H = 4,\,5$, and some $\mu_H$ above 5, respectively. As $\mu_H$ further increases, the flattened-out cross section starts to increase across all $\theta$. 
The reason for this is that when $\omega^2$ drops below $V_{\mathrm{eff}}(l=1,\mu_H)$, no $l \geq 1$ wave can enter the effective potential without much difficulty. In other words, at most the $s$ wave can experience some amount of absorption and nontrivial scattering by the effective potential, while all higher-$l$ partial waves are deflected to infinity. This results in flattened scattering cross sections, as seen for the $(\omega=1,\,\mu_H=3,4,5)$ and $(\omega=2,\,\mu_H=5)$ cases. This fact is reflected in Fig. \ref{fig:abs2}, in that when $\omega = 1,\,2$, respectively, only the $\mu_H$ up to 2 and 3 cases have nonzero absorption.  

Only when at least two low-$l$ partial waves can enter the potential barrier and be nontrivially scattered can interference between waves of different $l$ occur, and consequently the interference fringes in the cross section can appear. Indeed, even before the effective potential in each plot exceeds $\omega^2$ as $\mu_H$ increases, we observe that the interference fringes become wider, with stronger contrast between different maxima. These are all consequences of fewer partial waves participating in the constructive/destructive interference. On the other hand, the increase of the flattened cross section as $\mu_H$ further increases is simply a reflection of the fact that larger $\mu_H$ generally corresponds to a higher and wider barrier (see Fig. \ref{fig:V1}). In the low-frequency limit, both of these changes imply a more repulsive effective potential, which increases the differential scattering cross section \cite{Sakurai:QM3,Suzuki}.

	\begin{figure*}[htp!]
		\centering
		\includegraphics[width=0.33\textwidth]{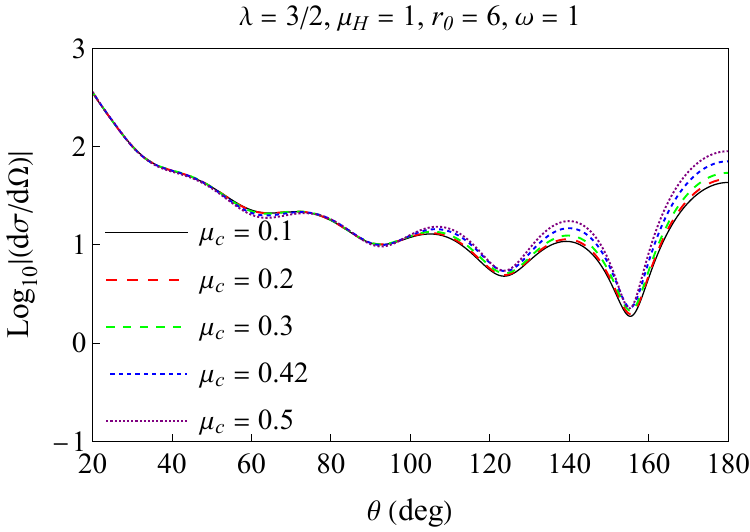}\includegraphics[width=0.33\textwidth]{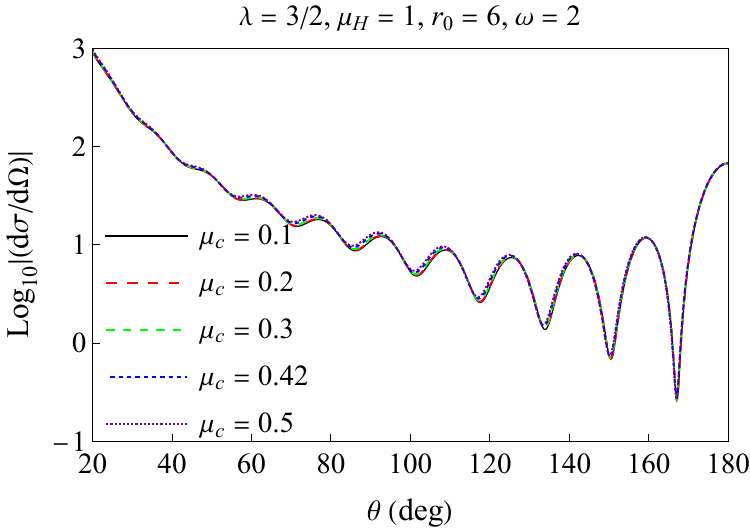}\includegraphics[width=0.33\textwidth]{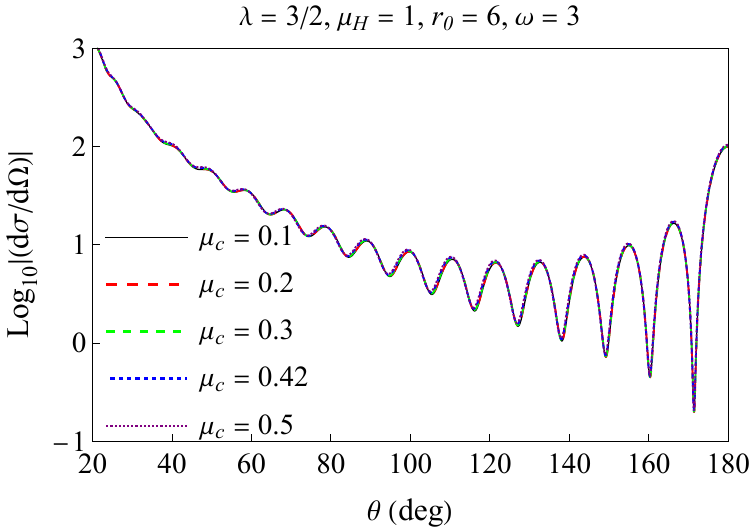}
		
		\includegraphics[width=0.33\textwidth]{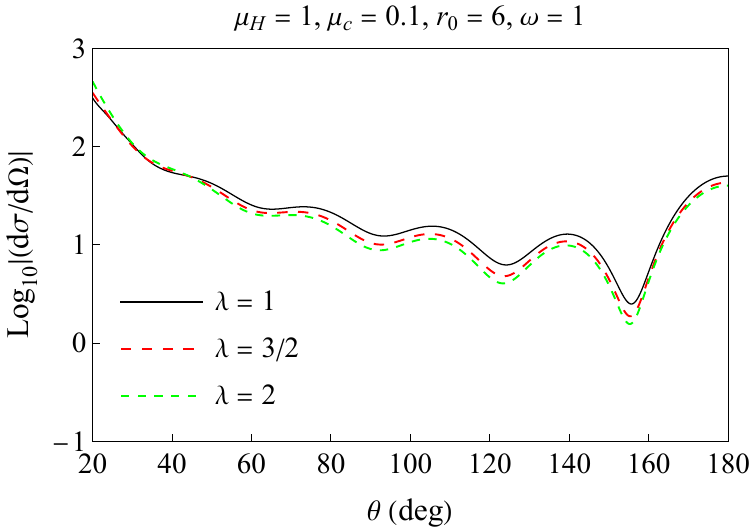}\includegraphics[width=0.33\textwidth]{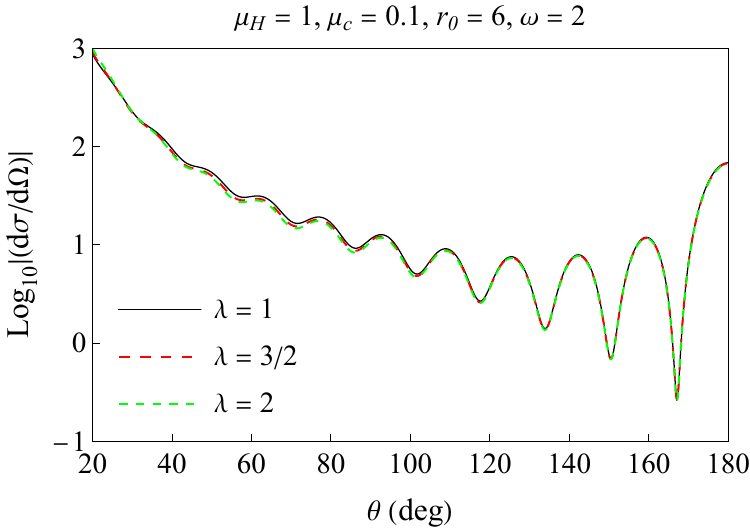}\includegraphics[width=0.33\textwidth]{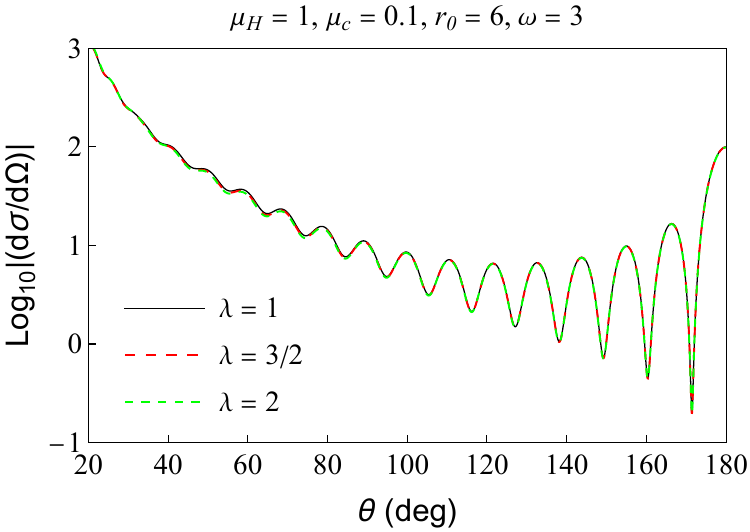}
		
		\includegraphics[width=0.33\textwidth]{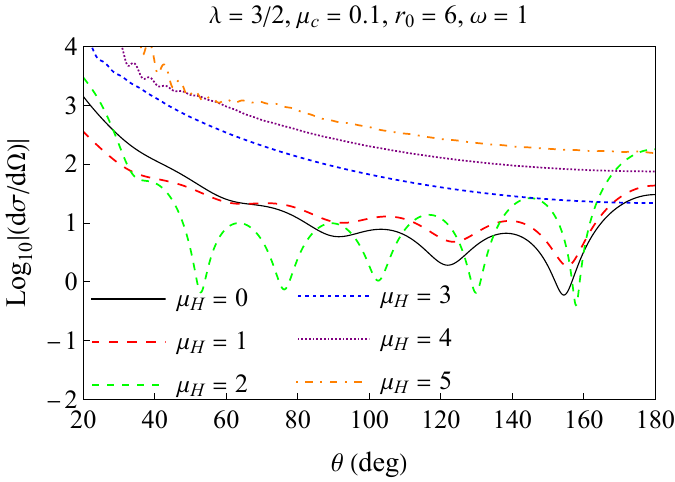}\includegraphics[width=0.33\textwidth]{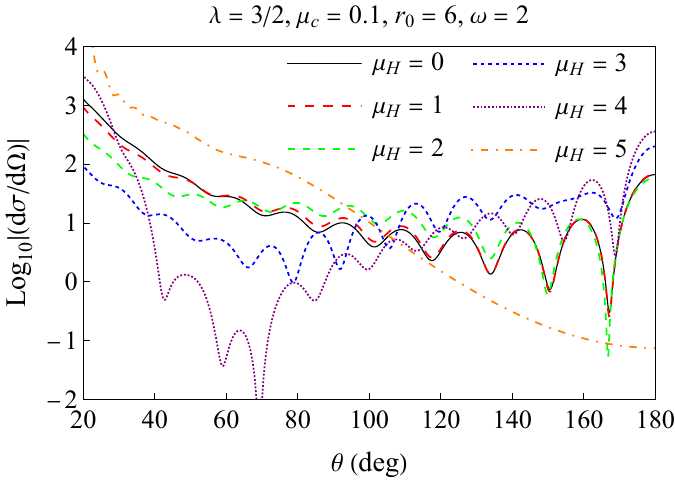}\includegraphics[width=0.33\textwidth]{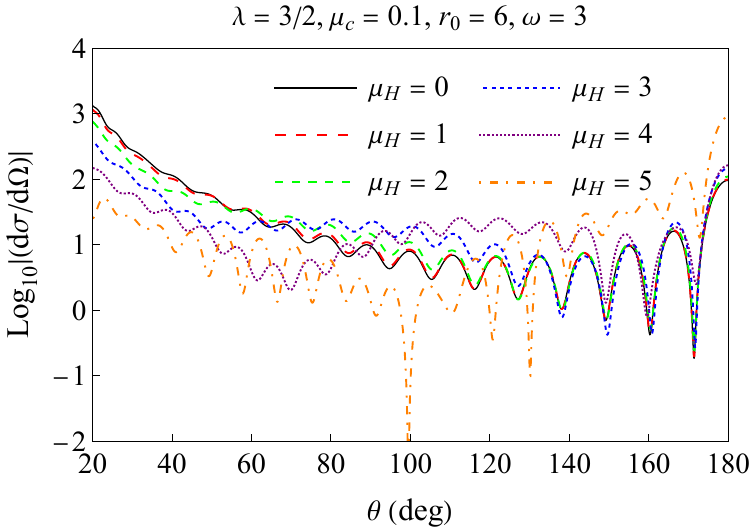}
		
		\includegraphics[width=0.33\textwidth]{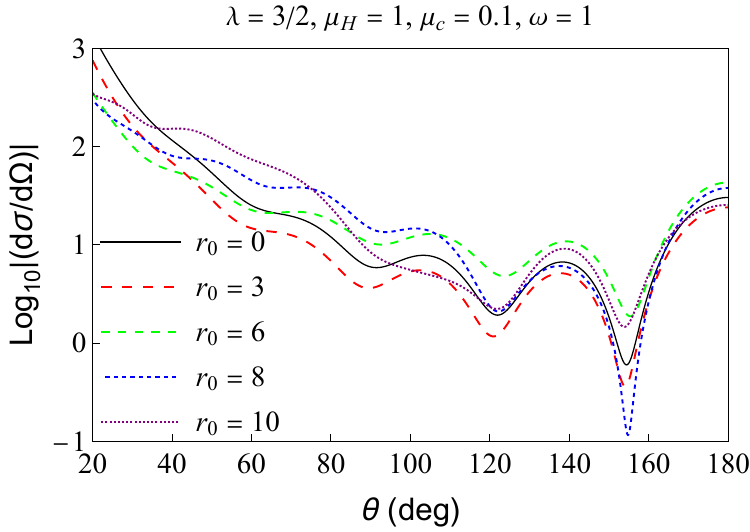}\includegraphics[width=0.33\textwidth]{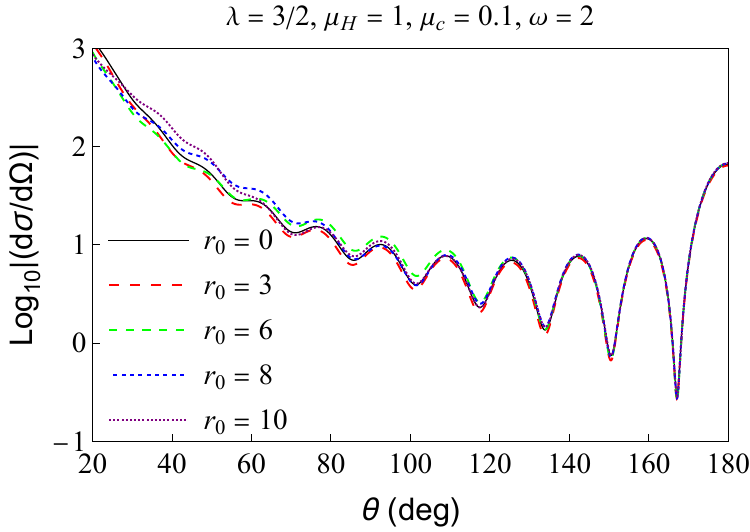}\includegraphics[width=0.33\textwidth]{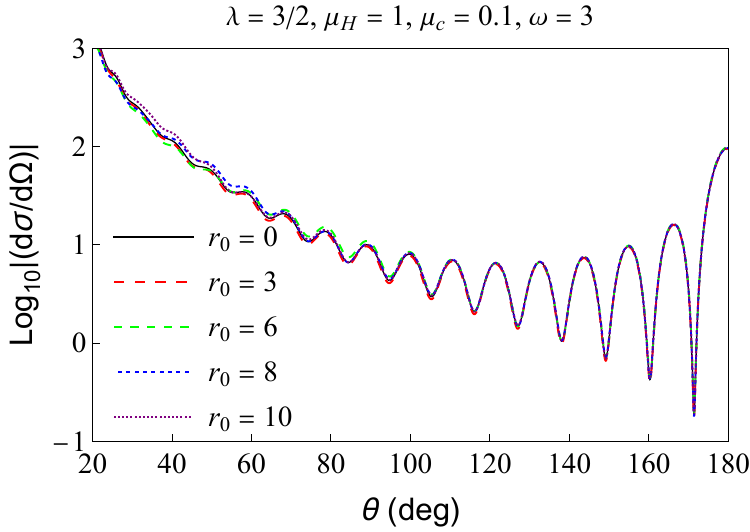}
		
		\caption{The differential scattering cross section in model~II for parameters $\mu_{c}$ (top), $\lambda$ (middle),  $\mu_{H}$ (bottom) and $r_0$ (extended-bottom)  at $\omega=1$ (left), $\omega=2$ (central) and $\omega=3$ (right).}
		\label{fig:sca2}
	\end{figure*}
	
In Fig. \ref{fig:sca2}, from the left to right columns, we show for model~II the differential scattering cross section of the incident waves with frequencies $\omega = 1,\,2,\,3$, respectively, with different parameters $\mu_{c}$, $\lambda$, $\mu_{H}$, and $r_0$ from the top to bottom rows. We find that, in addition to the common features observed in model~I, especially the general property that the effects of the parameters are more apparent for smaller $\omega$, the differential cross sections here show some unique features, which we will discuss next one by one.

The first feature is about the effect of the horizon mass $\mu_H$ (see the third row of Fig. \ref{fig:sca2}). Here we observe from the third-row plots that, besides the flattening of the scattering cross sections after $\mu_H$ surpasses certain values, which is similar to the case of model~I, the scattering cross sections exhibit some irregular dips. This is most obvious for the $(\omega=2,\, \mu_H=4)$ curve and the $(\omega=3,\,\mu_H=5)$ curve. These dips are also due to the existence of potential wells in the effective potential, as seen in Fig. \ref{fig:V2}. For example, for the $\omega=2$ case with $\mu_H=0,1,2,$ or $3$, it is found by analyzing the effective potential for these cases that some of the low-$l$ partial waves are able to enter the outer barrier, but most of them then either further enter or bounce back from the inner barrier.
On the other hand, if $\mu_H=5$ or above, then most partial waves, except for the  low-$l$ ones, will not even be able to enter the outer barrier. For the choices of $\mu_H$ in the plot, only $\mu_H=4$ allows some partial waves to experience nontrivial scattering between the two barriers around the potential well. Their scattering effectively produces rapid variations in the phase shifts, leading to pronounced cancellations among themselves and with other non-resonant partial waves in the scattering amplitude at some angles around $70^\circ$.
For the $\omega=3$ case, as seen in the right panel of the third row of Fig. \ref{fig:sca2}, the effective potential shows that basically it is the same group of $l$ (around 15 to 18) waves that can experience the resonance. According to the impact-parameter relation $b\sim (l+1/2)/\omega$, these partial waves are scattered with a closer approach to the center and experience a larger deflection angle. The corresponding dip then occurs at a larger scattering angle, which is around $100^\circ$ now. We note that a similar decrease in the scattering cross section when resonance occurs was also observed in scattering by wormholes \cite{LimaJunior:2022zvu}.

The second feature is about the effect of the parameters on the high-frequency scattering cross section at large scattering angles. We observe from the $\omega=3$ case in Fig. \ref{fig:sca2} that $(\mu_c, \lambda, r_0)$ have little effect on the cross section at $\theta\sim 180^\circ$, and for the parameter $\mu_H$, a similar effect can be seen for waves with even larger $\omega$ (not shown). To understand this, we note that in BH scattering, it is the interference between partial waves related to the unstable photon sphere, or equivalently the critical impact parameter $b_c$, that leads to the glory scattering in the backward direction \cite{Andersson:1995vi}. As shown in Fig. \ref{fig:abs2}, the optical limit of the total absorption cross section in the high-frequency regime, i.e., $\sigma_\mathrm{abs}\approx \pi b_c^2$, exhibits only a very weak dependence on the parameters $(\mu_c, \lambda, r_0)$. This indicates that these parameters have a negligible effect on the critical impact parameter $b_c$. Consequently, one expects that the glory scattering cross section at large $\theta$ for high-frequency waves should also display only a weak dependence on $(\mu_c, \lambda, r_0)$, which is exactly what is observed. Indeed, the above reasoning also applies to the effect of $\mu_c$ and $\lambda$ in model~I, and this feature is also present in the top-right and middle-right panels in Fig. \ref{fig:sca1}, although for our parameter ranges it is not as apparent as in Fig. \ref{fig:sca2} for model~II.

Finally, the parameter $r_0$ is unique to model~II. It is seen from the last row of Fig. \ref{fig:sca2} that its effect on the differential scattering cross section is also most apparent for low-energy waves, but this effect is quite complex, given that it can either increase or decrease the cross section depending on the scattering angle. The former feature has a similar explanation as for other parameters, namely that low-energy waves are more sensitive to the small potential-height changes caused by $r_0$, which can be seen from the lower-left panel of Fig. \ref{fig:V2}. The complexity of the effect of $r_0$ can be understood from the same potential plot. It is seen there that as $r_0$ changes, not only do the potential heights change, but also the (average) locations/radii of the potential wells drift. The former determines which and how many partial waves will effectively experience the potential, while the latter determines at what radius (or equivalently, scattering angle) the resonance will occur. These two main factors collectively determine the effect of $r_0$ in this plot.

	\section{Conclusions}\label{sec:conclusion}
	
In scalar–tensor theories, stationary and asymptotically flat BHs are indistinguishable from those in GR \cite{Sotiriou:2011dz,Berti:2015itd}. However, the presence of matter drastically alters the situation because the coupling between matter and the scalar field leads to an effective mass for the scalar field. In this work, we investigate the impact of this effective mass on Schwarzschild BH absorption and scattering, using two density-profile models of the effective mass introduced in Ref.~\cite{Dima:2020rzg}.  

In model~I, the absorption cross section increases as the mass parameter $\mu_c$ and the slope index $\lambda$ increase, or as the horizon mass $\mu_H$ decreases.
The presence of a nonzero mass $\mu_H$ significantly enhances the effective potential, preventing the divergence of the absorption cross section in the low-energy limit $\omega \to \mu_c$, reminiscent of charged BHs absorbing scalar waves with the same sign of charge \cite{Li:2024xyu}. In model~II, however, the parameters $(\mu_{c},\, \lambda,\, \mu_{H},\, r_0)$ in the range we considered mainly modify the low-frequency absorption behavior, and typically result in the appearance of one or two resonance peaks in the absorption cross section. At high frequencies, the absorption is almost unaffected by the parameters $(\lambda,\, \mu_{H},\, r_0)$, and the influence of the parameter $\mu_{c}$ agrees with that in model~I. This is because the parameters $(\mu_{c},\,\lambda,\, \mu_{H},\, r_0)$ mainly enhance the outer potential, while the parameters $(\mu_{c},\,l)$ govern the inner potential, which may form a potential well. Due to the fact that the potential well caused by the effective mass is asymmetric, the number of resonance peaks is significantly smaller than in the case of wormholes \cite{Lima:2020auu,Magalhaes:2023har,Furuta:2024jpw,Konoplya:2025hgp} or other exotic compact objects \cite{Macedo:2018yoi}.  

The differential scattering cross sections in both models exhibit some common features of BH scattering, such as forward divergence, backward glory, and enhanced oscillation amplitudes at larger scattering angles, all consistent with the results for massive scalar fields. Besides, in model~I, the increase of $\mu_c$ and $\lambda$ enhances the average scattering flux. The increase of $\mu_c$ and decrease of $\lambda$ also lead to broader fringes. These effects weaken at higher frequencies. In contrast, the horizon mass $\mu_H$ plays a more intricate role by substantially enhancing the potential. Its increase first suppresses the cross sections in general and widens all interference fringes, until it passes a critical value such that no $l\geq 1$ partial waves can effectively overcome the potential barrier, at which point the scattering cross section becomes flattened for all scattering angles and increases as $\mu_H$, and consequently the potential, increases.  

The presence of potential wells in model~II leads to additional scattering features beyond those observed in model~I. The effect of the parameter $\mu_H$ here is particularly interesting because the cross section shows a dip at intermediate scattering angles for certain choices of $(\omega,\,\mu_H)$, due to the appearance of the potential well for these parameters and the resonant scattering of certain partial waves around it. The effects of the parameters $\mu_c$ and $\lambda$ are also shown to diminish in the high-frequency and high-scattering-angle limit. Finally, the effect of the parameter $r_0$ is also most apparent for low-frequency scattering. Moreover, $r_0$ affects not only the potential height but also the potential well location and consequently causes a non-monotonic, angle-dependent effect on the differential cross section.

	\section*{Acknowledgements}
	
	This work is supported by the National Natural Science Foundation of China.

\end{document}